\documentclass[aps,pre,onecolumn,showpacs,showkeys,square,numbers,amssymb,amsmath,nobibnotes,nofootinbib]{revtex4}
\usepackage[utf8]{inputenc}%
\usepackage[T1]{fontenc}%
\usepackage{bm}%
\usepackage[colorlinks=true,linkcolor=red]{hyperref}
\usepackage{amsmath}
\usepackage{amsfonts}
\usepackage{amssymb}
\usepackage{verbatim}
\usepackage{epsfig}
\usepackage{graphicx}
\usepackage{soul}
\usepackage[sort&compress]{natbib}
\usepackage{times}

\newcommand{\beq}{\begin{equation}}
\newcommand{\eeq}{\end{equation}}
\newcommand{\beql}{\begin{eqnarray}}
\newcommand{\eeql}{\end{eqnarray}}

\newcommand{\lp}{\left(}
\newcommand{\rp}{\right)}

\newcommand{\nn}{\nonumber}
\newcommand{\om}{\omega}
\newcommand{\al}{\alpha}
\newcommand{\ga}{\gamma}
\newcommand{\lam}{\lambda}

\newcommand{\la}{\lambda}
\newcommand{\e}{\mbox{e}}

\newcommand{\cI}{\oint_{\cal C}\frac{d\om}{2\pi \mathrm{i}}}

\newcommand{\Vp}{V^{\prime}_m}
\begin{document}


\title{A singular-potential random matrix model arising in mean-field glassy systems\footnote{This paper is dedicated to the memory of Oriol Bohigas.}}



\author{Gernot Akemann$^1$, Dario Villamaina$^2$ and Pierpaolo Vivo$^3$}
\affiliation{$1.$ Fakult\"at f\"ur Physik, Universit\"at Bielefeld, Postfach 100131, D-33501 Bielefeld (Germany)\\$2.$ Laboratoire de Physique Th\'eorique de l'ENS \& Institut de
  Physique Th\'eorique Philippe Meyer,\\ 24 rue Lhomond 75005 Paris (France)\\$3.$ Laboratoire de Physique Th\'{e}orique et Mod\`{e}les
Statistiques (UMR 8626 du CNRS), Universit\'{e} Paris-Sud,
B\^{a}timent 100, 91405 Orsay Cedex (France)}



\date{\today}

\begin{abstract}
We consider an invariant random matrix model where the standard Gaussian potential is distorted by an additional single pole of order $m$. We compute the average or macroscopic spectral density in the limit of large matrix size, solving the loop equation with the additional constraint of vanishing trace on average. The density is generally supported on two disconnected intervals lying on the two sides of the pole. In the limit of having no pole, we recover the standard semicircle. Obtained in the planar limit, our results apply to matrices with orthogonal, unitary or symplectic symmetry alike. The orthogonal case with $m=2$ is motivated by an application to spin glass physics. In the Sherrington-Kirkpatrick mean-field model, in the paramagnetic phase and for sufficiently large systems the spin glass susceptibility is a random variable, depending on the realization of disorder. It is essentially given by a linear statistics on the eigenvalues of the coupling matrix. As such its large deviation function can be computed using standard Coulomb fluid techniques. The resulting free energy of the associated fluid precisely corresponds to the partition function of our random matrix model. Numerical simulations provide an excellent confirmation of our analytical results.

\end{abstract}

\pacs{02.10.Yn,02.50.-r,64.70.Q-}
\keywords{spin glass, Sherrington-Kirkpatrick model, random matrix theory, two-cut density, spin glass susceptibility.}

\maketitle




\section{Introduction}

Since their inception in nuclear physics more than sixty years ago, and even before in applied statistics, ensembles of matrices with random entries have found an impressive number of applications (see \cite{mehta,oxford,guhr,forrestergas} for a quite exhaustive account).
One of the applications of random matrix theory (RMT) has been to spin-glass physics, a field that has seen a spectacular growth in the past thirty years with a number of exciting and often counter-intuitive results \cite{mezard,zamponi}. One of the main features of spin glasses is the existence of a corrugated free energy landscape at low temperature, characterized by the presence of many minima that trap the dynamics for long time and break ergodicity. Given that the coupling between spins is through a random matrix, the statistics of stationary points of a random free energy landscape has attracted much interest in recent years using RMT tools \cite{bouc,fyod1,fyod2,auffinger,aging,braydean,fyodnadal,cavagnagiardina,cavagnaparisi,deanmaj}. RMT was also employed to model structural glasses in high dimensions \cite{cugliandolo, deo, parisiliquid} and universal RMT predictions were used as a reference point to study instantaneous normal modes in amorphous materials and liquids \cite{franz,pandey,mezardzee}.

In spite of these important but limited connections, it seems that the full power of RMT has not yet been exploited in the context of spin glass physics. The purpose of this paper is to prepare an exact computation (under very mild assumptions) of the distribution of the spin glass susceptibility in
the Sherrington-Kirkpatrick (SK) mean-field model. To reach this goal (which is detailed in Section \ref{sec:applications}) we need to analyze an invariant RMT where the confining potential is the sum of two terms: the standard Gaussian part, and a singular term consisting of a second order pole. More generally, we will consider an $m$-th order pole, and the potential thus reads
\begin{equation}
V_m(x)=\frac{1}{2}x^2+\frac{2A}{(x-a)^m}\ ,
\label{Vmdef1}
\end{equation}
with $a>0$ and 
$A\in\mathbb{R}$.
Random matrix ensembles with a singular potential and notably with poles have been already considered in the literature, see e.g. \cite{berry,brouwer,chen,mezzadri1,mezzadri2,majtexier,chi}. For example in \cite{mezzadri2} the microscopic limit of a modified Gaussian model including first and second order poles was considered, and a connection was found to the Painlev\'e III equation. A similar potential, this time in the chiral or Laguerre class, was introduced to study the transition from the Bessel to the Airy kernel \cite{chi}. The same potential also appears as a moment generating function in the problem of Wigner delay time in chaotic cavities \cite{majtexier}.

\begin{figure}[h!bt]
\begin{center}
\includegraphics[width=0.5\columnwidth,clip=true]{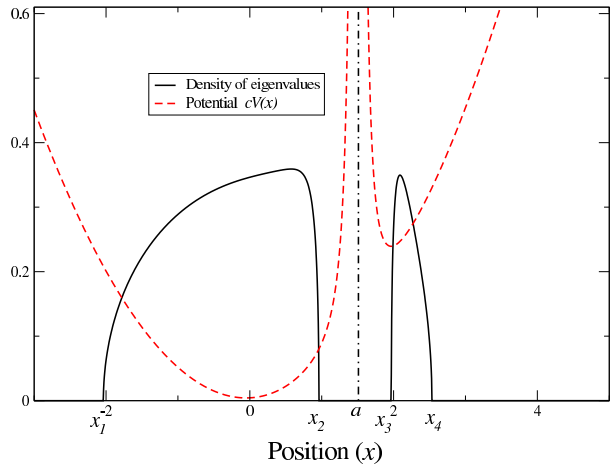}
\caption{Spectral density for a specific choice of the confining singular
  potential $V_2(x)$ (here $a=1.5$ and $A=0.1$). The density (solid black line) is
 supported on two disjoint intervals around the minima of the confining potential (red dashed line). The potential in the figure is rescaled by a factor $c=0.1$ for graphical reasons.}\label{fig:density1}
\end{center}
\end{figure}

Here we compute the average (macroscopic) spectral density of the RMT model defined by the potential \eqref{Vmdef1}, while in a forthcoming publication we will analyze the consequence of this calculation for the SK problem summarized in Section \ref{sec:applications}. The density $\rho_\star(\lambda)$ (see Eq. \eqref{density} for $m=2$) is generally supported on two disconnected intervals on the two sides of the pole at $\lambda=a$  (see Fig. \ref{fig:density1} for $m=2$ and $A=0.1$), unless $A=0$ or $a\to\infty$ where the solution becomes the standard semicircle.

The technical tools we use for our computation are the loop equations for the resolvent. It has cuts along the support of the density and is therefore called two-cut in our case. From the loop equations we obtain three equations for the four edge points of the support $\{x_1,x_2,x_3,x_4\}$. The fourth equation necessary to close the system is found by imposing that the ensemble is traceless on average, a condition that for $m=2$ is needed in the spin glass problem (see Section \ref{sec:applications} for details).

Is a two-cut solution the only possibility? The same calculation can be repeated assuming a one-cut solution instead, with edge points $\{y_1,y_2\}$. Without imposing the traceless constraint, we are led to two equations for $y_1$ and $y_2$ (see Appendix \ref{m2example1cut}), however the traceless case (relevant for the spin glass applications) makes the system overdetermined and does not lead to a consistent one-cut density anywhere in the $(A,a)$ plane, except for $A=0$ or $a\to\infty$.

Since our two-cut solution for the density is derived in the planar limit, it
applies equally well to all three symmetry classes (in particular to real symmetric matrices, relevant to the SK problem). This is in contrast with most other results listed above that are limited to the unitary symmetry class. The solution we present is of more mathematical interest for the RMT community, due to the technical complications arising from the two-cut nature of the resolvent:  it also turns out that the constraint of zero trace will play an important role in our calculation, both in fixing the parameters of the two-cut solution and in excluding a consistent one-cut solution which is not the semicircle.

The rest of the paper is organized as follows. In Section \ref{sec:applications} we provide a self-contained introduction to the physics of the SK model that motivates our study. First, we describe how the spin glass susceptibility (a standard indicator of the onset of a spin glass phase) at the paramagnetic minimum and for sufficiently large systems is a random variable, depending on the realization of disorder. It can be written as a linear statistics on the eigenvalue of the inverse susceptibility matrix. The latter is related in a simple way to the coupling matrix, which is drawn (using a very mild assumption) from the standard Gaussian Orthogonal Ensembles (GOE) ensemble. Phrasing the problem in terms of the distribution of a linear statistics on GOE eigenvalues, we can then use the Coulomb fluid technique to address its large deviation properties via the saddle point method (see subsection \ref{secsec:Coulomb}). The free energy of the associated Coulomb fluid is precisely linked to the partition function of the matrix model introduced here (see \eqref{Vmdef1} with $m=2$) for real symmetric traceless matrices. The density \eqref{density} (plotted in Fig. \ref{fig:density1}) is just the equilibrium density of the associated Coulomb fluid. At the end of Section \ref{sec:applications}, we will also present its explicit expression for those readers not interested in RMT technicalities.
Section \ref{sec:planar} contains the main calculations of this paper. Here we derive the planar loop equation for a singular potential and solve it for the resolvent with a general $m$th order pole in the two-cut situation. The explicit solution for $m=2$, relevant for the SK model, is then spelled out in great detail, and we also analyze the phase boundary for the two-cut solution. The putative one-cut solution is postponed to the Appendix \ref{m2example1cut}. This is because (as announced earlier) it turns out that for traceless matrices the one-cut solution is inconsistent, unless the pole disappears. In this case, the model is just Gaussian and hence its density is the semicircle.
In Section \ref{sec:numerical} we perform sophisticated numerical simulations to test our formula for the density and in Section \ref{sec:conclusions} we offer concluding remarks and perspectives for future work.

\setcounter{equation}{0}\section{Application to spin glasses and main result}\label{sec:applications}

\subsection{General setting}

We consider the Sherrington-Kirkpatrick (SK) model \cite{sherrington1975solvable}, a mean-field spin glass model defined by the Hamiltonian
\begin{equation} \label{3}
\mathcal{H} [ \{ S_i\} ,\{ x_{ij} \}]= \frac{-\,J}{2N^{1/2}} \sum_{i \neq j}^ {N}x_{ij}S_i S_j  + \sum_{i=1}^ {N} h_i S_i\ ,
\end{equation}
where $S_i = \pm 1$ are Ising spins and the all-to-all couplings $\{ x_{ij} \}_{i > j=1, \cdots, N} \equiv \{ x \},\, x_{ji} \equiv x_{ij} \forall i>j$ are distributed according to a standard normal distribution. Such couplings collectively define the \emph{quenched disorder} of the ensemble. This means that thermodynamical observables depending on the spin configurations are obtained by first averaging with respect to the Gibbs-Boltzmann (canonical) weight at inverse temperature $\beta$, and then averaging over the disorder (distribution of the $\{x\}$). The two different averages are denoted by $\langle (\cdots)\rangle$ and $\overline{(\cdots)}$ respectively. The strength of the disorder is tuned by the parameter $J$.

The celebrated Parisi solution \cite{parisi1980order,parisi1983order, talagrand2003generalized, mezard,0,NishimoriBook01} indicates that the SK model undergoes a spin-glass transition (in zero external fields) at the critical temperature $T_c = 1/\beta_c=J$ in the thermodynamic limit $N \rightarrow \infty$, where ergodicity breaking occurs and the spin-glass susceptibility defined below diverges \cite{mezard,NishimoriBook01,0}.

One way to understand this mechanism was originally proposed by Thouless, Anderson and Palmer (TAP) \cite{thouless1977solution}. The idea can be considered as a generalization of the Curie-Weiss approach to the ferromagnetic transition: since the SK model is fully connected, it lacks any spatial structure and in the thermodynamic limit the organization of the states is determined only by the local magnetizations $m_i$. Hence, in the TAP approach one writes the free energy $\mathcal{F}(\{ m_i \}, \beta) $ of the system as a function of fixed local magnetizations $m_i$, and studies the resulting free energy landscape. These local magnetizations $m_i$ are the canonical average $\left\langle\cdots \right\rangle$ of the spin $S_i$ performed with the Gibbs-Boltzmann weight at fixed disorder $\{ x \}$ and inverse temperature $\beta$.

\begin{figure*}[t]
\begin{centering}
\includegraphics[width=7cm]{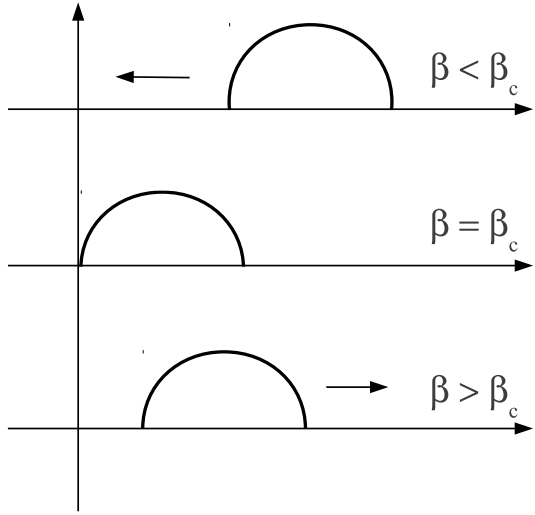}
\caption{Sketch of the behavior of the eigenvalue density of the inverse susceptibility matrix \eqref{eq:hes}.\label{fig11}  }
\end{centering}
\end{figure*}

The \emph{minima} of the free energy landscape are clearly crucial to characterize the phases of the system. One should distinguish the high-temperature ($\beta<\beta_c$) from the low temperature ($\beta>\beta_c$) phase: at $N \to \infty$ and high temperature the only minimum of $\beta \mathcal{F}(\{ m \},\beta)$ is the \emph{paramagnetic} one  with $m_i = 0$ $\forall i$. On the contrary, in the low temperature phase, the TAP free energy has exponentially many different minima, a typical signature of a \emph{glassy} phase, where the system is trapped for long time within minima of the landscape and ergodicity is broken. So, how does this TAP free energy look like? Plefka \cite{plefka1982convergence} showed that it can be obtained as an expansion in powers of the parameter $\alpha \equiv \frac{\beta J} { N^{1/2}}$ (high-temperature expansion), resulting in
\begin{equation} \label{3b}
 -\beta \mathcal{F}(\{ m_i \}, \beta) \simeq  - \sum_{i} \left[ \frac{1+m_i}{2} \ln \left(  \frac{1+m_i}{2} \right) + \frac{1-m_i}{2} \ln \left(  \frac{1-m_i}{2} \right) \right] +\frac{\alpha}{2}  \sum_{(ij)} x_{ij} m_i m_j  + \frac{ \alpha^2}{4}\sum_{(ij)} x_{ij}^2(1-m_i^2) (1-m_j^2),
\end{equation}
where $(ij)$ stands for the sum over all distinct pairs, and one retains only the first three terms: the first two are just the standard mean-field approximation, while the third one is called the Onsager reaction term. Further terms can be systematically included (Georges-Yedidia expansion \cite{yedidia1990fully}), but they vanish anyway for the SK model as $N\to\infty$, therefore they can be safely neglected. This expansion has been extensively used for several systems, both in the classical \cite{georges1990low,yedidia1990fully,yokota1995ordered}, and quantum domain \cite{plefka2006expansion, ishii1985effect, de1992cavity, biroli2001quantum}. The stability pattern of extremal points (maxima, minima and saddles) in this multidimensional free-energy landscape is encoded in the Hessian of $\mathcal{F}$ (or \emph{inverse susceptibility matrix})
\beq \label{73}
\beta \chi^ {-1}_{ij}  \equiv  \beta \frac{\partial h_i}{\partial m_j}  =  \frac{\partial^ 2 (\beta \mathcal{F})}{\partial m_i \partial m_j} \ ,
\eeq
which at the paramagnetic minimum $m_i=0$ reads from \eqref{3b} (to leading order\footnote{Equation \eqref{eq:hes} is obtained by replacing $x_{ij}^2$ with $\overline{x_{ij}^2}=1$ in the last term of \eqref{3b}. This is only correct to leading order in $N$ as it amounts to neglect fluctuations of the couplings altogether. For finite $N$, there is a correction term for the diagonal entries of \eqref{73}, see \cite{zarinelli2}, that correlates diagonal and off-diagonal elements. For sufficiently large systems, this correction leaves the ensemble traceless on average and does not significantly alter the spectral properties of a standard GOE matrix, therefore we safely ignore it.} in $N$)
 \begin{equation}\label{eq:hes}
\beta \chi^ {-1}_{ij} = (1 + \beta^2 J^2) \delta_{ij} - \alpha x_{ij}  \quad .
\end{equation}
Given that $\{x\}$ are random variables, the inverse susceptibility matrix \eqref{eq:hes} is a random matrix, whose spectrum gives information about the stability of the paramagnetic minimum. The standard (albeit heuristic) argument goes as follows. Given that the matrix $\{x\}$ belongs to the GOE ensemble with the extra constraint of having zeros on the diagonal, $x_{ii}=0$ from \eqref{3}, the average spectral density of $ \chi^ {-1}_{ij}$ is a shifted semicircle (see Fig. \ref{fig11}). At high temperature $(\beta<\beta_c)$, the spectrum of the Hessian has support on the positive region, therefore the paramagnetic minimum is stable. At $\beta=\beta_c$, the edge of the semicircle touches zero, signaling the appearance of zero modes and consequently the onset of an instability of the paramagnetic phase \cite{bray1979evidence}. However, for $\beta>\beta_c$ the semicircle comes back to the positive side, and therefore it seems that the paramagnet is stable at all temperatures. This result is in fact incorrect for $\beta>1/J$ \cite{giard} and the paramagnet becomes indeed unstable at low temperature. However, the picture in Fig. \ref{fig11} still suggests that the critical temperature is essentially related to the appearance of zero modes in the \emph{average} spectrum of the inverse susceptibility matrix (a shifted semicircle at $N\to\infty$).

To be more precise, a convenient measure (built upon the Hessian eigenvalues) to detect the onset of a spin-glass phase is the \emph{spin-glass susceptibility} $\chi_{\mathrm{SG}}^x(\beta,N)$, defined as
\begin{equation}
\chi_{\mathrm{SG}}^x(\beta,N)=\frac{1}{N}\mathrm{Tr}[ \chi_{ij}^2 ]\ ,
\label{defchi}
\end{equation}
where the susceptibility matrix at the paramagnetic minimum is defined in \eqref{eq:hes}. It is therefore a random variable (depending parametrically on the inverse temperature $\beta$ and system size $N$) which fluctuates from one realization of disorder to another. This is signaled by the superscript $^x$. It can be proven that such quantity is proportional to the square of the overlap between two sample at fixed disorder (see e.g. \cite{zamponi} and \cite{monthus,yucesoy,middleton} for recent numerical and analytical study on overlap distribution).

If we now average over the disorder, and define $\chi_{\mathrm{SG}}(\beta,N)=\overline{\chi_{\mathrm{SG}}^x(\beta,N)}$, this averaged susceptibility (still depending parametrically on $N$ and $\beta$) is a non-decreasing function of $\beta$ (see e.g. Fig. 1 in \cite{zarinelli1}) such that for $N\to\infty$, $\chi_{\mathrm{SG}}(\beta,N\to\infty)$ is finite in the paramagnetic region ($\beta<\beta_c$) and is divergent in the spin-glass phase ($\beta>\beta_c$). Due to this different behavior when crossing $\beta=\beta_c$ in the thermodynamic limit, this susceptibility is indeed a good indicator of the onset of a glassy phase.

What can be said about the fluctuations of $\chi_{\mathrm{SG}}^x(\beta,N)$ around its average value for large but finite $N$? Analytical arguments and numerical estimates \cite{zarinelli1,marinari,finitesize} yield a \emph{typical} scale of fluctuations of $\mathcal{O}(N^{-1/3})$, i.e. one writes
\begin{equation}
\chi_{\mathrm{SG}}^x(\beta,N)=\overline{\chi_{\mathrm{SG}}^x(\beta,N)}+N^{-1/3}\xi\ .
\end{equation}
Here the random variable $\xi$ has at this scale a limiting $N$-independent distribution
\begin{equation}
\lim_{N\to\infty}\mathrm{Prob}[\xi <z]=\mathrm{F}_\beta(z)\ .
\end{equation}
Note that such result is valid only in the paramagnetic phase and for system sizes so large that $\beta_c-\beta\gg N^{-1/3}$, otherwise the paramagnetic minimum, where \eqref{eq:hes} holds, may not be the relevant one. To the best of our knowledge, the limiting distribution $\mathrm{F}_\beta(z)$ is unknown to date. On the other hand, the random variable $\chi_{\mathrm{SG}}^x(\beta,N)$ also enjoys \emph{atypically} large (rare) fluctuations to the left and right of the mean, where the susceptibility takes values much smaller or larger than expected (see e.g. \cite{parisirizzo} for other studies of large deviations in the SK model). Such fluctuations are \emph{not} described by the scaling function $\mathrm{F}_\beta(z)$, but instead are governed by a large deviation function (see \cite{touchette} for an excellent review on large deviations), and in the next subsection we will describe a strategy based on the Coulomb fluid technique of RMT to compute it. The matching between the large deviation function close to the mean and the typical behavior on a scale of $\mathcal{O}(N^{-1/3})$ should also shed light on the tails of the scaling function $\mathrm{F}_\beta(z)$ itself, in complete analogy with what happens e.g. for the typical/atypical fluctuations of the largest eigenvalue of random matrices \cite{schehrnew} or the statistics of the ground state energy in disordered models \cite{monthusmatching}.

There is yet another interesting application of the calculation we prepare in the next subsection. Clearly, the sharp divergence of susceptibility, that can only happen at $N\to\infty$, is replaced by a smooth crossover for finite $N$. This leads to the (non-unique) definition of a \emph{pseudo-critical inverse temperature} as a random variable $\beta_{pc}^x(N)$ (depending on system size and realization of disorder) such that $\lim_{N\to\infty} \overline{\beta_{pc}^x(N)}=\beta_c$. This object in some sense marks the transition between a finite and a diverging susceptibility $\chi_{\mathrm{SG}}^x(\beta,N)$. What is the typical size of fluctuations with $N$ of $\beta_{pc}^x(N)$, and its limiting distribution as $N\to\infty$?

Two different groups \cite{zarinelli2,zarinelli1,marinari} have lately investigated these questions via extensive numerical simulations and analytical arguments, and two proposals for the limiting distribution (Gaussian or Tracy-Widom) were put forward. The main points of disagreement, summarized in Section IIIB of \cite{marinari}, seem mostly due to the choice of different algorithms to define the pseudo-critical inverse temperature. Whatever definition is used, however, the important point is that $\beta_{pc}^x(N)$ is a random variable precisely determined by the behavior of $\chi_{\mathrm{SG}}^x(\beta,N)$ as a function of $\beta$. As such, \emph{for a given definition}, its distribution is uniquely determined by the distribution of $\chi_{\mathrm{SG}}^x(\beta,N)$ itself, whose calculation in the large deviation regime is prepared here. Therefore we expect that such computation will eventually shed some light on the limiting distribution of $\beta_{pc}^x(N)$ as well. In the next subsection, we set up the computation of the distribution (in the large deviation regime) of spin-glass susceptibility as a RMT problem.

\subsection{Distribution of spin glass susceptibility of SK as a RMT problem}

We are now ready to prepare the computation of the large deviation function of the spin glass susceptibility defined in \eqref{defchi}. Hereafter we will set $J=1$ without loss of generality. In the TAP approximation in the paramagnetic phase, the inverse susceptibility matrix is given by \eqref{eq:hes}
 \begin{equation}\label{eq:hes2}
 \chi^ {-1}_{ij} = (\beta+\beta^{-1}) \delta_{ij} - \frac{1}{\sqrt{N}} x_{ij}  \quad ,
\end{equation}
where we have ignored the finite $N$ correction term multiplying $\beta$, and the coupling matrix $x_{ij}$ just belongs to the GOE, with the extra constraint $x_{ii}=0\quad\forall i$ on the diagonal. Random matrix models with constraints have been considered previously in the literature (see e.g. \cite{larsen,staring2003random, shukla2005random, bai2008large}). However, the presence of constraints on the entries could be potentially harmful, as it typically destroys rotational invariance. This would hinder the determination of the joint probability density of eigenvalues and therefore the exact solvability, a crucial ingredient for our calculation. On the other hand, it is known that zero-mean diagonal constraints as in this case are harmless for sufficiently large matrices: for example, the spectral density and the largest eigenvalue \cite{soshnikov1999universality} are virtually unaffected by it on average. Therefore, with the aim of retaining the exact solvability of the model, we simply draw the coupling matrix $x_{ij}$ from a (traceless) GOE. This is the only (very mild) assumption in an otherwise exact RMT approach.

Combining \eqref{eq:hes2} and \eqref{defchi}, the spin glass susceptibility $\chi_{SG}^x(\beta,N)$ is therefore a $\mathcal{O}(1)$ real random variable that can be written in terms of the rescaled eigenvalues $\{\lambda_i\}$ of $x_{ij}$ as:
\begin{equation}
\chi_{\mathrm{SG}}^x(\beta,N)=\frac{1}{N}\sum_{i=1}^N \frac{1}{(a-\lambda_i)^2}\ ,\label{chilinear}
\end{equation}
where
\beq
a\equiv \beta+\beta^{-1}.
\label{acrit}
\eeq
One may appreciate the divergence of $\chi_{\mathrm{SG}}^x(\beta,N)$ when the eigenvalues get close to the critical value $a=2$ (edge of the semicircle, see again Fig. \ref{fig11}).
In the paramagnetic phase ($\beta<\beta_c=1$, where \eqref{eq:hes} holds), $a> 2$. However, we will study the associated RMT problem in the more general setting $a\geq0$. For the standard GOE the spectral density $\rho_\star(\la)$ in the large-$N$ limit is the celebrated semi-circle, $\rho_\star(\la)=\frac{1}{2\pi}\sqrt{4-\la^2}$, and thus at the critical temperature, $a$ precisely hits the edge of the semicircle $(\la=2)$. Written in the form \eqref{chilinear}, the spin-glass susceptibility is a \emph{linear statistics}\footnote{A linear statistics is a random variable of the form $\phi=\sum_i f(\lambda_i)$, which does not contain products of different eigenvalues. The function $f(x)$ might well be highly non-linear, as it is in the present case.} on the eigenvalues of a (traceless) GOE matrix. Distributions of linear statistics on the eigenvalues of random matrices have been extensively studied both in physics \cite{politzer,chenlinear,forresterbaker,chenlinear2,vivoconductance,vivoindex} and mathematics (see e.g. \cite{soslinear} and references therein).

The eigenvalues $\{\lambda_i\}$ (assumed of $\mathcal{O}(1)$ for $N\to\infty$) of a zero-trace GOE random matrix are distributed according to the following joint law\footnote{Note that in the RMT literature the term \emph{fixed-trace ensembles} is usually employed when fixing the {\it second} moment to a constant non-zero value (see e.g. \cite{akemannfixed1}). After taking the large-$N$ limit in the unconstrained GOE, the first moment vanishes automatically.}
\begin{equation}
\mathcal{P}(\lambda_1,\ldots,\lambda_N):=\frac{1}{\mathcal{Z}_N}\e^{-\frac{N}{4}\sum_{i=1}^N \lambda_i^2}\prod_{j>k}^N|\lambda_j-\lambda_k|\ \delta\left(\sum_{i=1}^N \lambda_i\right) \ .
\end{equation}
Here the variance of the matrix elements is chosen in such a way that the limiting semi-circle for the spectral density extends between $[-2,2]$, and $\mathcal{Z}_N$ is a normalization constant.
Therefore the probability density of the spin-glass susceptibility $\chi_{\mathrm{SG}}^x(\beta,N)$
\begin{equation}
\mathcal{P}(\chi;a,N):=\mathrm{Prob}[\chi<\chi_{\mathrm{SG}}^x(\beta,N)<\chi+d\chi]
\end{equation}
(in the paramagnetic phase and for sufficiently large $N$) can be written as
\begin{equation}
\mathcal{P}(\chi;a,N)=\frac{1}{\mathcal{Z}_N}\int_{(-\infty,\infty)^N}d\lambda_1\cdots d\lambda_N \e^{-\frac{N}{4}\sum_{i=1}^N \lambda_i^2}\prod_{j>k}^N|\lambda_j-\lambda_k|\delta\left(\chi-\frac{1}{N}\sum_{i=1}^N \frac{1}{(a-\lambda_i)^2}\right)\ \delta\left(\sum_{i=1}^N\lambda_i\right)\ .
\label{main}
\end{equation}
Introducing an integral representation for the two delta functions, we obtain
\begin{equation}
\mathcal{P}(\chi;a,N)\propto \int \frac{dp}{2\pi}\e^{\mathrm{i}p\chi}
\int \frac{d\kappa}{2\pi}\int_{(-\infty,\infty)^N}d\lambda_1\cdots d\lambda_N \e^{-\frac{N}{4}\sum_{i=1}^N \lambda_i^2+\mathrm{i}\kappa\sum_{i=1}^N\lambda_i-\frac{\mathrm{i}p}{N}\sum_{i=1}^N \frac{1}{(a-\lambda_i)^2}}
\prod_{j>k}^N|\lambda_j-\lambda_k|
\label{main2}
\end{equation}
The $N$-fold $\{\lambda\}$ integral corresponds to the partition function of our singular-potential random matrix model\footnote{In the loop equation approach we will not impose the zero trace constraint by a delta function, hence there will be no linear term in the confining potential.}, see \eqref{Vmdef1} for $m=2$. Eq. \eqref{main2} for large $N$ is well-suited to a large deviation treatment based on the Coulomb fluid method, originally popularized by Dyson \cite{dyson} and recently employed in many different problems (see e.g. \cite{schehrnew} and references therein). In the next subsection, we will prepare this Coulomb fluid treatment, which will highlight the importance of the average spectral density of this model in the determination of the large deviation tails of the susceptibility.

\subsection{Coulomb fluid formulation and saddle point analysis}\label{secsec:Coulomb}

We will now take the large-$N$ limit and perform a saddle point analysis of the $N$-fold integrals form the previous subsection.
Exponentiating the Vandermonde determinant and introducing a continuum density of eigenvalues
\begin{equation}
\label{rhodef}
\rho(\lambda)=N^{-1}\sum_{i=1}^N \delta(\lambda-\lambda_i)\ ,
\end{equation}
we can replace sums with integrals using the rule
\begin{equation}
\sum_i g(\lambda_i)\to N\int d\lambda\rho(\lambda)g(\lambda)
\end{equation}
and suitably renaming $p$ and $\kappa$ we get:
\begin{equation}
\mathcal{P}(\chi;a,N)\propto \int d A dB d C\int\mathcal{D}[\rho]\exp\left\{-N^2 \mathcal{S}[\rho]
+\mathcal{O}(N)\right\}\ .\label{SP}
\end{equation}
Here the continuum \emph{action} $\mathcal{S}$ (depending parametrically on the Lagrange multipliers $A,B,C$ and on $a$ and $\chi$) is given by
\begin{align}
\nonumber \mathcal{S}[\rho] =&\frac{1}{4}\int d\lambda \lambda^2 \rho(\lambda)-\frac{1}{2}\iint d\lambda d\lambda^\prime \rho(\lambda)\rho(\lambda^\prime)\ln |\lambda-\lambda^\prime|
+A\left(\int d\lambda\frac{\rho(\lambda)}{(a-\lambda)^2}-\chi\right)\\
&+B\left(\int d\lambda \rho(\lambda)-1\right)+C\left(\int d\lambda\ \lambda\rho(\lambda)\right)\label{actionS}\ ,
\end{align}
where $B$ is an extra Lagrange multiplier enforcing the normalization of the density to unity. Written in the form \eqref{SP}, the probability density $\mathcal{P}(\chi;a,N)$ is just the canonical partition function at inverse temperature\footnote{Take care in distinguishing the inverse temperature $\beta_D$ of the auxiliary Coulomb fluid from the inverse temperature of the SK model $\beta$, at which the spin glass susceptibility is evaluated.} $\beta_D=1$ of an associated fluid of many particles in equilibrium under competing interactions: a confining single-particle potential (Gaussian plus second-order pole) and a repulsive all-to-all logarithmic potential. The action $\mathcal{S}$ is just the leading $N$ contribution to the free energy of the fluid, whose equilibrium density $\rho_\star$ is computed below using a saddle-point method. Evaluating the action at the saddle point, we get that the probability density of the spin glass susceptibility decays for large $N$ as
\begin{equation}
\mathcal{P}(\chi;a,N)\approx \exp\left[- N^2 \psi(\chi;a)\right],
\end{equation}
where the \emph{large deviation function} $\psi(\chi;a)$, supported on $\chi\in (0,\infty)$, is just given by
\begin{equation}
\psi(\chi;a) = \mathcal{S}[\rho_\star]-\mathcal{S}[\rho_\star]\Big|_{A\to 0}\ .
\end{equation}
Here we subtracted the asymptotic contribution coming from the normalization constant $\mathcal{Z}_N$,
and $\approx$ stands for a logarithmic equivalence $\lim_{N\to\infty} -\ln \mathcal{P}(\chi;a,N) /N^2=\psi(\chi;a)$. On general grounds, we expect the rate function to be a convex function, with a zero $\psi(\chi_0;a)=0$ at the average value of the susceptibility at $N\to\infty$, i.e.
\begin{equation}
\chi_0=\lim_{N\to\infty}\overline{\chi_{\mathrm{SG}}^x(\beta\ll 1,N)}=\int_{-2}^2 dx\frac{\sqrt{4-x^2}}{ 2 \pi (a-x)^2}=\frac{\Theta(a-2)}{2} \left(\frac{a}{\sqrt{a^2-4}}-1\right)=\frac{\beta^2}{1-\beta^2}
\Theta(1-\beta)\ ,
\label{chinot}
\end{equation}
where $\Theta(x)$ is the Heaviside function.
The last equality is obtained restoring the definition $a=\beta+\beta^{-1}$ (see \eqref{acrit}).
Note once again that the physical requirement $a>2$ is imposed by the condition $\beta\ll\beta_c=1$ that the system is in the paramagnetic phase, or equivalently the susceptibility is far from the diverging point $\beta=\beta_c=1$ (see \eqref{chinot}). However, in the following we will study the more general case $a\geq0$ and a pole of order $m$ in the single particle potential.

As anticipated, Eq. \eqref{SP} is amenable to a saddle point evaluation. Taking derivatives with respect to $\rho$, $A$, $B$ and $C$ we obtain:
\begin{align}
\frac{\delta \mathcal{S}}{\delta\rho} &= \frac{1}{4}\lambda^2-\int d\lambda^\prime \rho(\lambda^\prime)\log |\lambda-\lambda^\prime|+\frac{A}{(a-\lambda)^2}+B+C\lambda\ ,\label{sp1}\\
\frac{\partial \mathcal{S}}{\partial A} &=\int d\lambda \frac{\rho(\lambda)}{(a-\lambda)^2}-\chi\ ,\\
\frac{\partial \mathcal{S}}{\partial B} &=\int d\lambda \rho(\lambda)-1\ ,\\
\frac{\partial \mathcal{S}}{\partial C} &=\int d\lambda\ \lambda\rho(\lambda)\ .
\label{tricomi1}
\end{align}
Equating these derivatives to zero, we get from \eqref{sp1}:
\begin{equation}
\int d\lambda^\prime \rho_\star(\lambda^\prime)\log |\lambda-\lambda^\prime|=\frac{\lambda^2}{4}+\frac{A}{(a-\lambda)^2}+B+C\lambda\ ,\label{firsteq}
\end{equation}
while the other equations enforce constraints on the equilibrium density $\rho_\star(\lambda)$, which will be a parametric function of $\chi$ and $a$.
After differentiating both sides of the equation with respect to $\lambda$ in \eqref{firsteq} we get:
\begin{equation}
\mathrm{Pr}\int d\lambda^\prime \frac{\rho_\star(\lambda^\prime)}{\lambda-\lambda^\prime}=\frac{\lambda}{2}-\frac{2A}{(\lambda-a)^3}+C\equiv\frac12 V_{2}(\la)^\prime +C\ ,
\label{singint}
\end{equation}
where $\mathrm{Pr}$ denotes Cauchy's principal part. The right hand side of the equation is now precisely identified with the derivative of our modified singular potential $V_m(\lambda)$ (see \eqref{Vmdef1}) with a second order pole ($m=2$).
The solution $\rho_\star(\lambda)$ of the singular integral equation \eqref{singint} then has to be supplemented with the constraints of normalization, zero trace, and fixed susceptibility, respectively,
\begin{align}
\int d\lambda \rho_\star(\lambda) &=1\ ,\label{co1}\\
\int d\lambda\ \lambda \rho_\star(\lambda) &=0\ ,\label{co2}\\
\int d\lambda \frac{\rho_\star(\lambda)}{(a-\lambda)^2} &=\chi\ ,\label{co3}
\end{align}
where the integrals run over the support of the density (itself yet to be determined). Finding the solution of the singular integral equation \eqref{singint} satisfying the constraints is the main technical challenge. The rest of the paper is devoted to the computation of this $\rho_\star(\lambda)$, using the loop equation technique. In particular we will from now on consider the second order pole stemming from the Lagrange multiplier $A$
as part of the potential,
\begin{equation}
V_2(x)=\frac{1}{2}x^2+\frac{2A}{(x-a)^2}\ ,
\label{V2def}
\end{equation}
or more generally
\begin{equation}
V_m(x)=\frac{1}{2}x^2+\frac{2A}{(x-a)^m}
\label{Vmdef}
\end{equation}
where $a\geq0$ and $A\in\mathbb{R}$.
In the following the zero trace constraint will not be imposed from the beginning by using a Lagrange multiplier, but later on using a condition emerging from the asymptotic expansion of the planar resolvent. For that reason we set $C=0$ below.

In some limiting cases, e.g. $a=0$, the saddle point equation \eqref{singint} can be solved using simpler techniques \cite{tricomi}. We will use such limiting cases as a check of our loop equation calculation. In the limiting cases $A\to 0$ or $a\to\infty$, our general solution becomes the standard semicircle.
In a forthcoming publication, we will discuss the physical implications of our result by computing the full action $\mathcal{S}[\rho_\star]$ at the saddle point and comparing it with data. In the next subsection, we briefly summarize our main result from Section \ref{sec:planar}.

\subsection{Main result}

We are now ready to present our main result. The equilibrium density $\rho_\star(\lambda)$ satisfying the singular integral equation
\begin{equation}
\mathrm{Pr}\int d\lambda^\prime \frac{\rho_\star(\lambda^\prime)}{\lambda-\lambda^\prime}=\frac12 V_{m=2}(\la)^\prime\ ,
\label{singint2}
\end{equation}
with the constraints \eqref{co1} and \eqref{co2} reads
\begin{equation}
\rho_\star(\lambda) = \frac{(\lambda^2+\al_1 \lambda
+\al_0)}{2\pi|\lambda-a|^3}\sqrt{(\lambda-x_1)(\lambda-x_2)(\lambda-x_3)(x_4-\lambda)}\ ,\quad\lambda\in\sigma
\label{density}
\end{equation}
where the endpoints of the support $\sigma=[x_1,x_2]\cup[x_3,x_4]$ are functions of $a\geq0$ and $A\geq 0$, determined by eqs \eqref{x1}, \eqref{x2}, \eqref{x3} and \eqref{x4}. The parameter $A$ in turn has to be determined as a function of $\chi$ from the constraint \eqref{co3}. Note that no stable solution for the density exists for $A<0$, and the requirement of zero trace prevents the existence of a single-cut phase everywhere in the $(A,a)$ plane.

For the general $m$ case, the density is given by
\beq
\rho_\star(\la)\ =\ \frac{1}{2\pi}|M_m(\la)|
\sqrt{(\la-x_1)(\la-x_2)(\la-x_3)(x_4-\la)}\ ,\ \ \la\in\sigma\ ,
\label{rhoinitial}
\eeq
with
\beq
M_m(p)\ \equiv\
 \frac{1}{(p-a)^{m+1}}\sum_{j=0}^m\al_jp^j\ \ .
\label{MPdefinitial}
\eeq
The coefficients $\alpha_j$ are determined by matching the coefficients in the expansion of \eqref{Mcontour2}, while the endpoints of the support are determined by equations \eqref{Opbis}, \eqref{O1bis}, \eqref{Opminus1} and \eqref{asymp-condbis}. Obviously for $m=2$, we recover \eqref{density}.
An example of the spectral density was already shown in Fig. \ref{fig:density1} for illustration, together with the corresponding potential for $m=2$.

\begin{figure}[h!bt]
\begin{center}
\includegraphics[width=0.4\columnwidth,clip=true]{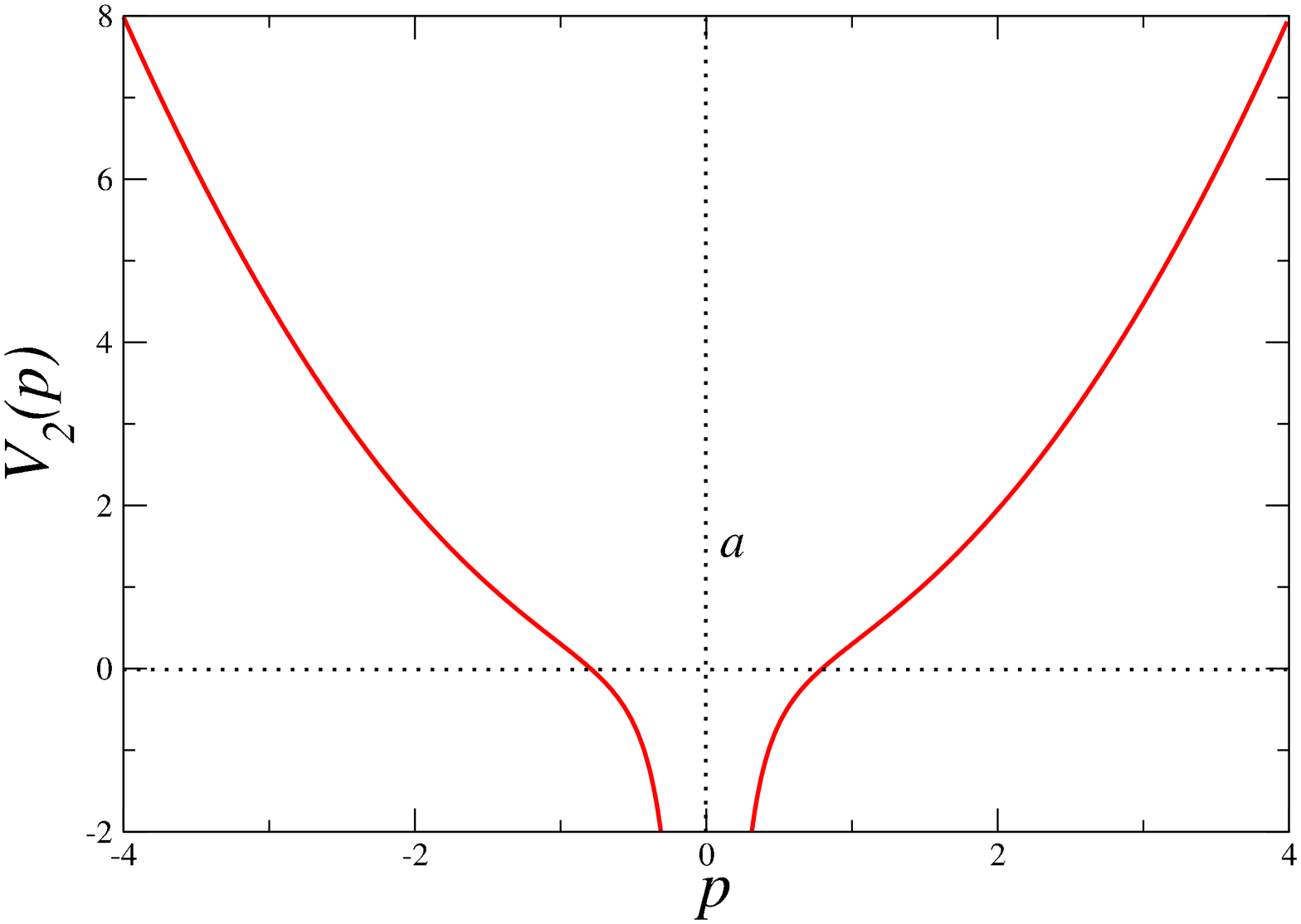}
\includegraphics[width=0.4\columnwidth,clip=true]{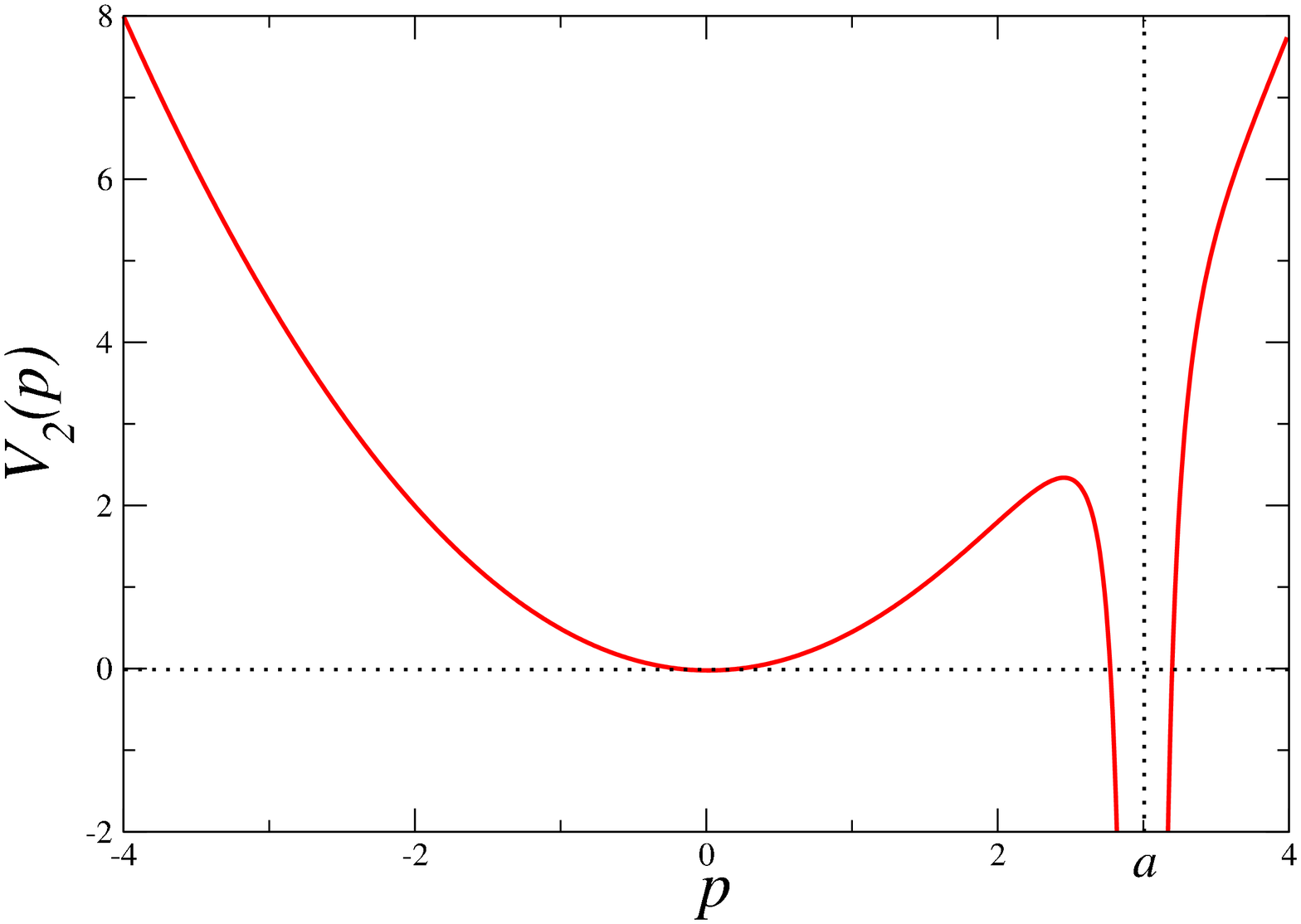}
\caption{\label{fig:density2} Singular potential $V_{m=2}(p)$ with negative $A=-0.1$: $a=0$ (left) and $a=3$ (right).}
\end{center}
\end{figure}

Why is a single support (one-cut) solution not stable in presence of the trace zero constraint?
Looking at Fig. \ref{fig:density1} for $A>0$, it is intuitively clear that the eigenvalues will favorably fill the two minima of the potential, rather than just a single minimum. Let us look now at the situation for a potential with $A<0$ depicted in Fig. \ref{fig:density2}. For $A<0$ and $a=0$ the potential is symmetric and no stable solution exists due to the unboundedness of the potential.  Moving the pole to the right, a single minimum develops which in the large-$N$ limit could in principle support a (metastable) single interval solution $\rho_\star(\la)$ (without imposing the traceless constraint). However, due to the pole which is now attractive the solution is always imbalanced, with $\int d\la \la \rho_\star(\la)>0$, and hence no one-cut solution exists for traceless matrices. This argument is of course sketchy and will be made more precise in the following Sections\footnote{Note that the support is not only determined by the potential minima but also by the effective interaction felt by an eigenvalue due to the $N-1$ surrounding ones.}.

\setcounter{equation}{0}

\section{Loop equation with singular potential in the planar limit}\label{sec:planar}

In the first subsection, we provide the planar limit of the
loop equation in the setting where our potential has a pole of order $m$ and the support $\sigma$ of the limiting eigenvalue density
$\rho_\star(\la)$ is composed of two disjoint intervals, separated by the pole.
This is the situation we expect from the previous discussion, and we will denote this two-cut solution by $\rho_\star^{(2)}(\la)$ to distinguish it from a putative one-cut solution $\rho_\star^{(1)}(\la)$ to be discussed later on.

In the second subsection, we construct
the two-cut solution for the planar resolvent, and the resulting density
$\rho_\star^{(2)}(\la)$ is obtained for the potential with a generic $m$th order pole. In the next subsection, this solution is most explicitly spelled our for the case of $m=2$, which we need for our application to the SK model.
Finally in the last subsection we determine the phase boundary of the two-cut phase.
The Ansatz for a putative one-cut solution is discussed in Appendix \ref{m2example1cut}.

\subsection{Loop equation for the planar resolvent}

The partition function of the our matrix model is
defined as
\beq
Z_N \equiv \
\int d\phi \ \mbox{exp}\left[-N \frac{\beta_D}{2} \mbox{Tr} V_m(\phi)\right] \ , \label{Z}
\eeq
where the integration measure $d\phi$ is either over the independent matrix elements of real symmetric, complex hermitian or quaternion self-dual
$N\times N$ matrices $\phi$. These three cases are labelled by the Dyson index $\beta_D=1,2,4$, respectively. However, when considering the planar limit $N\to\infty$ this distinction will become immaterial.
The matrix potential
\beq
V_m(\phi) \ \equiv \ \frac{1}{2} \phi^2+\frac{2A}{(\phi-a)^m} 
\eeq
has a pole of order $m$, $a\geq0$ and $A\in\mathbb{R}$ are real parameters. Note that in contrast with the previous Section, no constraint has been imposed on the eigenvalues $\la_{i=1,\ldots,N}$ of the matrix $\phi$ so far.
Averages are defined as usual by
\beq
\langle Q(\phi) \rangle \ = \ \frac{1}{Z_N} \int d\phi \ Q(\phi)\
                      \mbox{exp}\left[-N \frac{\beta_D}{2} \mbox{Tr} V_m(\phi)\right]\ .
\eeq
The basic object of our study is the resolvent or moment generating
function defined as
\beql
W(p) \ &\equiv& \  \frac{1}{N} \left\langle \mbox{Tr} \frac{1}{p-\phi}
\right\rangle
  \ = \ \frac{1}{N} \sum_{k=0}^\infty
             \frac{\langle \mbox{Tr}\phi^k \rangle}{p^{k+1}},\qquad p\in\mathbb{C} \setminus \sigma
   \ . \label{W}
\eeql
We will also need the connected $(conn)$ two-point resolvent defined as
\beq
W(p,q) \ \equiv \
\left\langle \mbox{Tr} \frac{1}{p-\phi}
 \mbox{Tr} \frac{1}{q-\phi}\right\rangle_{conn}
\equiv\left\langle \mbox{Tr} \frac{1}{p-\phi}
 \mbox{Tr} \frac{1}{q-\phi}\right\rangle -\left\langle \mbox{Tr} \frac{1}{p-\phi} \right\rangle
\left\langle \mbox{Tr} \frac{1}{q-\phi}\right\rangle
\ ,
\eeq
to formulate the loop equation for the resolvent. Here again $p$ and $q$ are complex variables outside the support $\sigma$ of the density.

In general both resolvents have a genus expansion in powers $1/N^g$ for $\beta_D=1,4$ and in powers of $1/N^{2g}$ for
$\beta_D=2$, where $g=0,1,2,\ldots$. These higher order terms are in principle to be determined by the loop equation, Eq. \eqref{loop} below. 
However, in the multi-cut case the situation is complicated due to additional correction terms that depend quasi-periodically on $N$. This is due to the discreteness of the eigenvalues, as was pointed out in \cite{BDE} (see also \cite{Deift}). They first enter in the connected two-point resolvent and are absent in the density to leading order.

Below, we will only be interested in the planar resolvent, the leading contribution in the large-$N$ limit:
\beq
\lim_{N\to\infty}W(p)\ \equiv \ W_0(p)+{\cal
  O}\left(\frac{1}{N}\right)
\ . \label{Wasympt}
\eeq
The leading asymptotic behavior for $W(p)$ and $W_0(p)$ for large $p$
is the same and follows from \eqref{W}:
\beq
\lim_{|p|\to\infty}W_0(p) \ \sim \ \frac{1}{p}\ +\ \left( \lim_{N\to\infty}\frac1N\langle \mbox{Tr}\phi \rangle\right)
\frac{1}{p^2}\ +\ {\cal  O}\left(\frac{1}{p^3}\right) \ . \label{W-pasympt}
\eeq
Here we have also displayed the second order term in the asymptotic expansion in $p$. If we impose the constraint of average zero trace of the matrix $\phi$ (relevant for the application to the SK model), then this term of order $1/p^2$ will have to vanish. We will come back to this later.

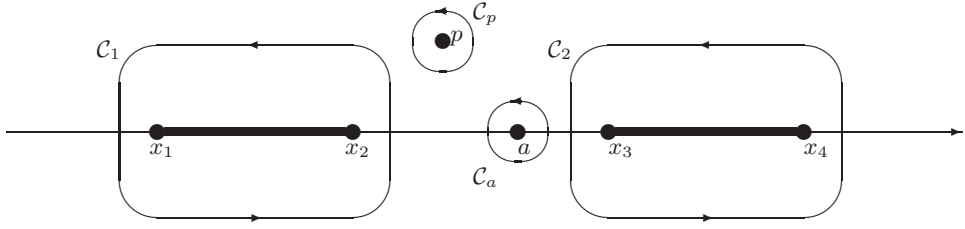
\begin{figure}[h]
\unitlength1cm
\begin{picture}(12.2,5.5)
\linethickness{1.0mm}
\multiput(2.1,2.3)(6,0){2}{\line(1,0){2.4}}
\thinlines
\put(0,2.3){\line(1,0){1.9}}
\put(4.7,2.3){\line(1,0){1}}
\put(6.9,2.3){\line(1,0){1}}
\put(10.7,2.3){\vector(1,0){2}}
\multiput(2,2.3)(2.6,0){2}{\circle*{0.2}}
\multiput(8,2.3)(2.6,0){2}{\circle*{0.2}}
\put(5.8,3.5){\circle*{0.2}}
\put(5.9,3.5){$p$}
\multiput(5.8,3.5)(1,-1.2){2}{\oval(0.8,0.8)[l]}
\multiput(5.8,3.5)(1,-1.2){2}{\oval(0.8,0.8)[r]}
\multiput(5.8,3.9)(1,-1.2){2}{\vector(-1,0){0.1}}
\put(5.7,2.3){\line(1,0){1}}
\put(6.8,2.3){\circle*{0.2}}
\put(6.8,2){$a$}
\put(6.2,1.6){${\cal C}_{a}$}
\put(6.2,3.8){${\cal C}_{p}$}
\put(1.9,2){$x_{1}$}
\put(4.5,2){$x_{2}$}
\put(8,2){$x_3$}
\put(10.6,2){$x_4$}
\multiput(2.,2.3)(6,0){2}{\oval(1.0,2.3)[l]}
\multiput(4.6,2.3)(6,0){2}{\oval(1.0,2.3)[r]}
\multiput(2.,3.45)(6,0){2}{\line(1,0){1.2}}
\multiput(3.4,1.15)(6,0){2}{\line(1,0){1.2}}
\multiput(4.6,3.45)(6,0){2}{\vector(-1,0){1.4}}
\multiput(2.,1.15)(6,0){2}{\vector(1,0){1.4}}
\put(1.2,3.3){${\cal C}_{1}$}
\put(7.2,3.3){${\cal C}_{2}$}
\end{picture}
\caption{The contour of integration
         $ {\cal C}={\cal C}_1\cup{\cal C}_2$
with respect to the location of the pole of the
  potential at $z=a$ and the argument of the resolvent at $z=p$.
The latter two are enclosed by ${\cal C}_a$ and ${\cal C}_p$, respectively. Note that we always have $p,a\notin\sigma$.}
                     \label{contour}
\end{figure}

The derivation of the loop equation for a multiple-interval support $\sigma$ of the limiting spectral density goes along the same lines as in \cite{AKE96} for $\beta_D=2$, and its extension to $\beta_D=1,4$ \cite{Itoi96},
exploiting the invariance of the partition function under a
field redefinition\footnote{Note that apart from the additional pole our definition of the potential differs from the one in \cite{Itoi96} by a prefactor of $\beta_D/2$. Also we have suppressed the quasi-periodic contributions from \cite{BDE} here.} $\phi \rightarrow \phi + \epsilon/(p-\phi)$:
\beq
W(p)^2 \  -\
\cI \frac{ V_m^{\prime}(\om)}{p-\om} W_0(\om)
\ =\ \frac1N \left(\frac{2}{\beta_D}-1\right)\frac{\partial}{\partial p}W(p)-\frac{1}{N^2}W(p,p)\ .\label{loop}
\eeq
In the planar limit we only keep the leading order terms on the left hand side, and we obtain
\beq
W_0(p)^2 \ = \
\cI \frac{ V_m^{\prime}(\om)}{p-\om} W_0(\om) \ ,\ \ p\not\in \sigma \ ,\label{planarloop}
\eeq
where in our case
\beq
V_m^{\prime}(w)=w-\frac{2mA}{(w-a)^{m+1}}\ .
\label{Vmprime}
\eeq
The $\beta_D$ dependence has dropped out here, and results for $\beta_D=2$ and $\beta_D=1,4$ differ only in the
the next correction which is of order $1/N^2$ or $1/N$, respectively.
Here and in the rest of this Section we assume a two-cut solution, as will become more clear in the next subsection. For the putative one-cut solution we refer to Appendix \ref{m2example1cut}.
The corresponding contour of
integration for two cuts ${\cal C}={\cal C}_1\cup{\cal C}_2$ in Eqs. \eqref{loop} and \eqref{planarloop}
is depicted in Fig. \ref{contour} enclosing the corresponding two-interval support
\beq
\sigma\ \equiv\ [x_{1},x_{2}]\cup[x_{3},x_{4}] \ ,
\ \ x_1< x_2<a<x_3<x_4 \ \ .
\eeq
Neither the argument of the planar resolvent $p$ on the right hand side of Eq. \eqref{planarloop}, nor the pole of the potential at $a$ are contained inside the integration contour ${\cal C}$, and hereafter we will always assume $p\neq a$.
Moreover, we also assume that $a\notin\sigma$. For $A>0$ this clearly cannot happen due to the repulsion of the potential whereas for $A<0$ this would lead to an instability because of the unboundedness of the potential.

We finally note the functional relation between the limiting macroscopic spectral density $\rho_\star(\la)$ and the
planar resolvent (valid for any number of cuts):
\beq
W_0(p) \ =\ \int_\sigma d\lam \frac{\rho_\star(\lam)}{p-\lam} \ ,
\ \ p\not\in\sigma\ .
\label{Wrho}
\eeq
It simply follows from the definition Eq. \eqref{W} by going to the eigenvalue representation and replacing the sum by an integral.
Below we will see that $W_0(p)$ has square root cuts along the support
$\sigma$, hence also the name two-cut case for our setup. The singular integral equation \ref{Wrho} can be inverted and the density reconstructed from $W_0(p)$ by taking the discontinuity along the cuts,
\beql
\rho_\star(\lam)\ &=& \ \frac{1}{2\pi \mathrm{i}}\lim_{\epsilon\to 0}
         \Big( W_0(\lam-\mathrm{i}\epsilon)-W_0(\lam+\mathrm{i}\epsilon) \Big)
                 \ ,\ \ \ \ \ \ \lam \in \sigma \ . \label{rhoW0}
\eeql

\subsection{The two-cut solution for a general pole of order $m$}\label{planar}

\indent

Equation \eqref{planarloop} for the planar resolvent $W_0(p)$
can be solved by mapping it to a quadratic equation. Deforming the contour in Eq.
\eqref{planarloop} to infinity one can exploit the asymptotic
behavior in Eq. \eqref{W-pasympt}, $W_0(p)\sim1/p$. In contrast with the standard multi-cut case
with non-singular potentials \cite{AKE96}, here the deformed contour picks up
an additional $m$-th order pole from the potential at $z=a$, as can be seen in Fig. \ref{contour}. One gets
\beq
(W_0(p))^2 \ = \ \Vp(p) W_0(p)+\frac{1}{m!}
\left.\left(\frac{2A}{(p-w)}W_0(w)\right)^{\!(m)}\right|_{w=a}
 +\oint_{\cal C_{\infty}}\frac{d\om}{2\pi \mathrm{i}} \frac{\Vp(\om)}{p-\om}W_0(\om) \ ,
\eeq
for the contributions from the poles at $p$, at $a$ and at $\infty$, respectively. Here the superscript $^{(m)}$ denotes the $m$-th derivative.
At infinity due to $W_0(p)\sim1/p$ only the Gaussian part of the potential contributes, and
we get as the final answer
\beq
(W_0(p))^2 \ =
\ \Vp(p)
W_0(p)+\frac{1}{m!}
\left.\left(\frac{2A}{(p-w)}W_0(w)\right)^{\!(m)}\right|_{w=a}
\ + 1.
\label{looe_4}
\eeq
Since the second term on the right hand side only depends on $W_0(a)$ and derivatives thereof, which are constant with respect to $p$, this equation is quadratic in $W_0(p)$. Its solution can be formally written
as
\beql
W_0(p) &=& \frac{1}{2}\Vp(p) \pm \frac{1}{2}\sqrt{(\Vp(p))^2+4Q(p)} \ ,
\nonumber\\
Q(p)   &=& \frac{1}{m!} \left.\left(\frac{2A}{(p-w)}W_0(w)\right)^{\!(m)}\right|_{w=a}
\ + 1\ .
\label{W0quad}
\eeql
While the rational function $Q(p)$ still implicitly depends on $W_0(a)$, this formal solution can be simplified.
Namely our assumption that $W_0(p)$ has $2$ square root cuts in the complex
plane leads to the following Ansatz:
\beq
W_0(p) \ \equiv\ \frac{1}{2}
\left(\Vp(p)-M_m(p)\sqrt{\prod\nolimits_{i=1}^{4}(p-x_i)}\right)\ ,
\label{ansatz}
\eeq
where
\beq
M_m(p)\ \equiv\
\frac{P_m(p)}{(p-a)^{m+1}}\ = \ \frac{1}{(p-a)^{m+1}}\sum_{j=0}^m\al_jp^j\ \ ,
\label{MPdef}
\eeq
is a rational function. Here the solution with the minus sign in front of the square root together with the choice of
branch of the square roots $\sqrt{\prod\nolimits_{i=1}^{4}(p-x_i)}\ \sim p^2$ for large $|p|\gg1$ is made to comply with the asymptotic behavior Eq. \eqref{W-pasympt}.
The fact that the polynomial $P_m(p)$ is of order $m$ follows from Eq. \eqref{W0quad}, upon bringing all terms in Eq. \eqref{Vmprime} on a common denominator and counting powers. We postpone the determination of the $m+1$ coefficients $\al_j$ and of the 4 endpoints of the support $x_l$ in terms of
the parameters of the potential $a$ and $ A$ because the expression for the rational function $M_m(p)$ and hence for the planar resolvent can be simplified. We only note at this stage that following Eq. \eqref{rhoW0} the Ansatz Eq. \eqref{ansatz} completely determines the spectral density\footnote{We have put an absolute value around the rational function here because the discontinuity in Eq. \eqref{rhoW0} has opposite signs along the two different cuts.}:
\beq
\rho_\star^{(2)}(\la)\ =\ \frac{1}{2\pi}|M_m(\la)|
\sqrt{(\la-x_1)(\la-x_2)(\la-x_3)(x_4-\la)}\ ,\ \ \la\in\sigma\ .
\label{rho}
\eeq
The rational function
$M_m(p)=P_m(p)/(p-a)^{m+1}$ can be written as a contour integral, being analytic everywhere except at $p=a$.
Denoting by ${\cal C}_p$ and by ${\cal C}_a$
the contours around $w=p$ and $w=a$ in the complex plane, see
Fig. \ref{contour}, we have
\beq
M_m(p)=\oint_{{\cal C}_p}\frac{dw}{2\pi
 \mathrm{i}}\frac{1}{w-p}M_m(w)=-
\oint_{{\cal C}_a}\frac{dw}{2\pi
  \mathrm{i}}\frac{1}{w-p}M_m(w)\ .
\label{Mcontour}
\eeq
This is because pulling the contour around $w=p$ to infinity will only give a contribution from
$w=a$
as $M_m(p)$ is analytic on $\sigma$, and the contribution at infinity
vanishes because of $M_m(p)\sim 1/p$ for large $p$. On the other hand we can solve
Eq. \eqref{ansatz} for  $M_m(p)$ and insert this into the integral in
Eq. \eqref{Mcontour}:
\beql
M_m(p)&=&
-\oint_{{\cal C}_a}\frac{dw}{2\pi \mathrm{i}}\frac{1}{w-p}
\frac{\Vp(w)-2W_0(w)}{\sqrt{\prod\nolimits_{i=1}^{4}(w-x_i)}}\nn\\
&=&
-\oint_{{\cal C}_a}\frac{dw}{2\pi  \mathrm{i}}
\frac{\Vp(w)}{(w-p)\sqrt{\prod\nolimits_{i=1}^{4}(w-x_i)}}\nn\\
&=&\oint_{{\cal C}_a}\frac{dw}{2\pi  \mathrm{i}}
\frac{2mA}{(w-p)(w-a)^{m+1}\sqrt{\prod\nolimits_{i=1}^{4}(w-x_i)}}
=\frac{1}{m!} \left.\left(\frac{2mA}{(w-p)\sqrt{\prod\nolimits_{i=1}^{4}(w-x_i)}}
\right)^{\!(m)}\right|_{w=a}\!.
\label{Mcontour2}
\eeql
In the first step we have assumed that $W_0(w)$ has no pole at $w=a$, which we will confirm self-consistently below, and hence its contribution vanishes. In the second step we have only kept the singular part of $\Vp(w)$. This form expresses the function $M_m(p)$ exclusively in terms of the 4 endpoints of the cuts $x_j$, which still remain to be determined.

With this result we may also simplify Eq. \eqref{ansatz} for $W_0(p)$. Writing the first
term there as a contour integral around $w=p$, and inserting the second line of Eq. \eqref{Mcontour2} into the second term in Eq. \eqref{ansatz} we have
\beql
W_0(p) &=&  \frac12\oint_{{\cal C}_p}\frac{dw}{2\pi \mathrm{i}}\frac{\Vp(w)}{w-p}
           \sqrt{\prod_{i=1}^{4}\lp \frac{p-x_i}{w-x_i}\rp}
           +\frac12 \oint_{{\cal C}_a}\frac{dw}{2\pi  \mathrm{i}}\frac{1}{w-p}
\frac{\Vp(w)}{\sqrt{\prod\nolimits_{i=1}^{4}(w-x_i)}}
\sqrt{\prod\nolimits_{i=1}^{4}(p-x_i)}\nn\\
&=& \frac{1}{2}\oint_{{\cal C}}\frac{dw}{2\pi \mathrm{i}}\frac{\Vp(w)}{p-w}
           \sqrt{\prod_{i=1}^{4}\lp \frac{p-x_i}{w-x_i}\rp} \ .\label{W0}
\eeql
Connecting the contours ${\cal C}_p$ and ${\cal C}_a$ and pulling it over the cuts to infinity, where the
contribution at infinity vanishes, leads to the second equation ($\cal C$ being the contour around both cuts). This is the standard form of the planar resolvent for multiple cuts as it was found in \cite{AKE96} for non-singular potentials. As a last step one can easily convince oneself that the limit $\lim_{p\to a}W_0(p)$ is non-singular,  being a rational function in $(a-x_i)$ with poles at the endpoints.
Hence our assumption that $W_0(p)$ does not have a pole in $p=a$ is self-consistent.
An explicit example for $W_0(a)$ will be given in the next subsection for $m=2$.

In order to complete our solution for the planar resolvent $W_0(p)$ in Eq. \eqref{W0} and hence for the limiting density $\rho_\star(\la)$ in Eq. \eqref{rho} we still need to determine the four endpoints $x_j$ of the support in
terms of the parameters of the potential $a$ and $A$.
Furthermore we also introduced the $m+1$ auxiliary coefficients $\al_j$ in Eq. \eqref{MPdef}, that parametrize the rational function $M_m(\la)$ inside the density. These coefficients $\al_j$ easily follow as functions of $a,A,m$ and the $x_{j=1,2,3,4}$ by comparing coefficients in Eq. \eqref{MPdef} and Eq. \eqref{Mcontour2}, and we will give an example for the $\al_j$ for $m=2$ below.

How can we determine the endpoints $x_j$ of the support? So far we have not yet used the asymptotic expansion Eq. \eqref{W-pasympt}, that the solution for $W_0(p)$ Eq. \eqref{W0}, or better Eq. \eqref{ansatz} has to satisfy.
In order to expand the latter for large $|p|\gg1$ we introduce the more convenient elementary symmetric functions $e_1,e_2,e_3,e_4$ of the variables $x_i$ as new variables,
\begin{equation}
e_1 =\sum_{i=1}^4 x_i\ ,\quad e_2 =\sum_{i<j}x_i x_j\ ,\quad e_3 =\sum_{i<j<k} x_i x_j x_k\ ,\quad e_4 = \prod_{i=1}^4 x_i\ .
\label{elem}
\end{equation}
This leads to
\beq
\prod_{i=1}^4(p-x_i)\ =\ p^4-p^3e_1+p^2e_2-pe_3+e_4\ \equiv\ F(p)\ ,
\label{Fdef}
\eeq
where we have introduced an abbreviation for this frequently appearing product.
This results in the expansion for $|p|\gg1$
\begin{equation}
\frac{\sqrt{F(p)}}{(p-a)^{m+1}}\approx \sum_{k=0}^3 \frac{c_k}{p^{k+m-1}}+\mathcal{O}\left(\frac{1}{p^{m+3}}\right)\ ,
\end{equation}
where
\begin{align}
c_0 &=1\ ,\\
c_1 &= (m+1)a-\frac{e_1}{2}\ ,\\
c_2 &= -\frac{m+1}{2}a e_1-\frac{e_1^2}{8}+\frac{e_2}{2}+\frac{(m+1)(m+2)}{2}a^2\ ,\\
c_3 &=\frac{\frac{8(m+1)(m+2)(m+3)}{3}a^3-[(2m+3)^2-1]a^2 e_1-(2m+2)a e_1^2-e_1^3+4 e_1 e_2 -8 e_3 +8(m+1)a e_2}{16}\ .
\end{align}
Put together with Eq. \eqref{MPdef} retaining only terms up to $\mathcal{O}(p^{-2})$ we have
\beql
\nn M_m(p)\sqrt{F(p)}&=&
\left(\sum_{j=0}^m \alpha_j p^j\right)
\left(\sum_{k=0}^3 \frac{c_k}{p^{k+m-1}}+\mathcal{O}\left(\frac{1}{p^{m+3}}\right) \right)\\
&\approx& \alpha_m p +\alpha_{m-1}+\alpha_m c_1+\frac{\alpha_{m-2}+\alpha_{m-1}c_1+\alpha_m c_2}{p}+\frac{\alpha_{m-3}+\alpha_{m-2}c_1+\alpha_{m-1} c_2+\alpha_m c_3}{p^2}+\mathcal{O}\left(\frac{1}{p^3}\right).\nn \\
&&
\eeql
Together with the expansion of the potential
\begin{equation}
V_m^\prime(p)\approx  p  - \frac{2A\delta_{m,1}}{p^2}+\mathcal{O}\left(\frac{1}{p^3}\right)
\end{equation}
we obtain the following three equations for the first three orders in the asymptotic expansion of $W_0(p)$ for large $p$ from Eq. \eqref{ansatz}, for arbitrary $m$:
\beql
{\cal O}(p):\ \ 0&=& 1-\al_m\ ,
\label{Opbis}\\
{\cal O}(1):\ \ 0&=& -\al_{m-1}-\al_m\left( (m+1)a-\frac12 e_1\right)\ ,
\label{O1bis}\\
\nonumber {\cal O}(p^{-1}):\ \ 1&=& -\frac12\left[\alpha_{m-2}+\alpha_{m-1}\left((m+1)a-\frac{e_1}{2}\right)+\alpha_m \left(-\frac{m+1}{2}a e_1-\frac{e_1^2}{8}+\frac{e_2}{2}+\frac{(m+1)(m+2)}{2}a^2\right)\right]
\ .\\ \label{Opminus1}
\eeql
Coefficients with negative index are defined to vanish, $\al_{-j}=0$ for $j=1,2,\ldots$.
After computing the $\al_{j=m,m-1,m-2}$ from \eqref{Mcontour2} in terms of the $e_{i=1,2,3,4}$ and the parameters of the potential we have three equations to determine the four unknowns $e_i$ (or equivalently the $x_i$).

This under-determination of the endpoints of the cuts is a well-known problem in the multi-cut solution of RMT \cite{Jurek}, and the number of missing equations increases with the number of cuts. 
There are several options to fix a meaningful planar limit. In \cite{Jurek} it was proposed to require equilibrium of chemical potentials among neighboring cuts. The idea was to allow for equilibration through eigenvalue tunneling at finite-$N$. However, due to the infinite potential barrier in our case such a prescription is not reasonable. A second option is to fix the filling fractions of eigenvalues on each interval of the support, see e.g. in \cite{BDE}. This would leave us with a single fraction for two cuts as a free parameter. 

Fortunately, in view of the application described in the previous Section \ref{sec:applications} we have a third option at hand. The constraint of a traceless matrix there is equivalent to the requirement that also the coefficient of order $1/p^2$ of the asymptotic expansion for large $|p|\gg1$, Eq. \eqref{W-pasympt}, of the resolvent vanishes:
\begin{align}
\nonumber {\cal O}(p^{-2}):\ \ 0&= \alpha_{m-3}+\alpha_{m-2}\left((m+1)a-\frac{e_1}{2}\right)+\alpha_{m-1} \left(-\frac{m+1}{2}a e_1-\frac{e_1^2}{8}+\frac{e_2}{2}+\frac{(m+1)(m+2)}{2}a^2\right)\\
\nonumber &+\alpha_m \left(\frac{\frac{8(m+1)(m+2)(m+3)}{3}a^3-[(2m+3)^2-1]a^2 e_1-(2m+2)a e_1^2-e_1^3+4 e_1 e_2 -8 e_3 +8(m+1)a e_2}{16}\right)\\
&- 2A\delta_{m,1}\label{asymp-condbis}
\end{align}
This gives the fourth equation needed to fix the endpoints completely. In the next Section, we explicitly give all details for the case of a second order pole $m=2$ as motivated by the application to the SK model.

\subsection{Explicit solution for the case $m=2$ }\label{m2example}

\indent

In this Section we write out explicitly the solution for the density including all its coefficients for the case $m=2$ which is relevant for the SK model from Section \ref{sec:applications}. Its potential is given by
\beq
V_2(x)=\frac{1}{2}x^2+\frac{2A}{(x-a)^2}\ ,
\label{V2def2}
\eeq
with the solution for the density reading
\beq
\rho_\star^{(2)}(\la)\ =\ \frac{1}{2\pi}|M_2(\la)|
\sqrt{(\la-x_1)(\la-x_2)(\la-x_3)(x_4-\la)}\ ,\ \ \la\in\sigma\ .
\label{rho2}
\eeq
Following Eq. \eqref{Mcontour2} we have for the rational function $M_2(p)$ that multiplies the square roots
\beql
M_2(p)&=&\left.\left(\frac{2A}{(w-p)\sqrt{F(w)
}}
\right)^{\prime\prime}\right|_{w=a}\nn\\
&=&\frac{-\,4A}{(a-p)^3F(a)^{1/2}}-\frac{2AF'(a)}{(a-p)^2F(a)^{3/2}}
+\frac{AF''(a)}{(a-p)F(a)^{3/2}}-\frac{3AF'(a)^2}{2(a-p)F(a)^{5/2}}
\label{M2}
\eeql
in terms of $F(a)$ from Eq. \eqref{Fdef}
and its first and second derivatives. Note that due to our choice of sign for the branch cut of $\sqrt{F(w)}\sim w^2$ at $|w|\to\infty$,
for $w=a$ in between the cuts this function is negative. However, in order to make our notation more suggestive we denote by the power $1/2$ the principal branch, $F(a)^{1/2}\equiv -\sqrt{F(a)}>0$, whereas the symbol $\sqrt{\hspace{1em}}$ denotes the function in the complex plane with the aforementioned choice of branch.

On the other hand we had defined the polynomial in the numerator of $M_2(p)$ to be
\beq
M_2(p)=\frac{\al_2p^2+\al_1p+\al_0}{(p-a)^3}
\ .
\label{M2P2}
\eeq
We can simply read off the coefficients from \eqref{M2} to be given by
\beql
\al_2&=&\frac{-A}{2F(a)^{5/2}}(2F(a)F''(a)-3 F'(a)^2)\ ,
\label{al2}\\
\al_1&=&\frac{-A}{F(a)^{5/2}}(2F(a)F'(a)-2aF(a)F''(a)+3aF'(a)^2)
\ ,
\label{al1}\\
\al_0&=&\frac{+A}{F(a)^{5/2}}
\left(4F(a)^2+2aF(a)F'(a)-a^2F(a)F''(a)+\frac32a^2F'(a)^2\right)
\ .
\label{al0}
\eeql
For completeness we also give the corresponding resolvent,
\beq
W_0(p)=\frac12\left(p-\frac{4A}{(p-a)^3} -M_2(p)\sqrt{F(p)}\right)\ .\label{corrresolvent}
\eeq
It is straightforward to check using Taylor expansion that the resolvent is non-singular in $p=a$:
\beq
W_0(a)=\frac{a}{2}+\frac{AF'''(a)}{6F(a)}-\frac{3AF'(a)F''(a)}{4F^2(a)}
+\frac{A(F'(a))^3}{F^3(a)}\ .
\label{W0ap}
\eeq

What remains to be determined are the positions of the 4 endpoints $x_{i=1,2,3,4}$ as functions of $a,A$. These are given by the asymptotic expansion of the planar resolvent, Eqs. \eqref{Opbis} - \eqref{Opminus1} for $m=2$,
\beql
\al_2&=& 1\ ,
\label{al2p}\\
\al_1&=&\frac{1}{2}e_1-3a\ ,
\label{al1p}\\
\al_0&=& 3a^2-2-\frac{3}{2}a e_1+\frac38e_1^2-\frac12e_2\ ,
\label{al0p}
\eeql
after inserting the expressions for the $\al_j$ from \eqref{al2} - \eqref{al0}. The fourth equation is given by Eq. \eqref{asymp-condbis} for $m=2$,
\beql
0&=&
\al_0 \left(3a-\frac{1}{2}e_1\right)
+\al_1\left(6a^2-\frac32 ae_1-\frac18 e_1^2+\frac12 e_2\right)
+\frac{\al_2}{16}\Big(160a^3-48a^2e_1-6ae_1^2-e_1^3+4e_1e_2-8e_3+24ae_2).
\nn\\ \label{fourth}
\eeql
Spelling these equations out most explicitly we have 

\beql
1&=&\frac{-A}{2F(a)^{5/2}}\big(2F(a)F''(a)-3 F'(a)^2\big)\ ,
\label{x1}\\
\frac{1}{2}e_1-3a&=&\frac{-A}{F(a)^{5/2}}\big(2F(a)F'(a)-2aF(a)F''(a)+3aF'(a)^2\big)
\ ,
\label{x2}\\
 3a^2-2-\frac{3}{2}a e_1+\frac38e_1^2-\frac12e_2&=&\frac{A}{F(a)^{5/2}}
\left(4F(a)^2+2aF(a)F'(a)-a^2F(a)F''(a)+\frac32a^2F'(a)^2\right)
\ ,
\label{x3}\\
0&=&
a^3-6a-\frac32 ae_2+\frac34ae_1^2-\frac32a^2e_1+e_1+\frac34 e_1e_2-\frac{5}{16}e_1^3-\frac12 e_3\ .
\label{x4}
\eeql
\begin{figure}[h!bt]
\begin{center}
\includegraphics[width=0.5\columnwidth,clip=true]{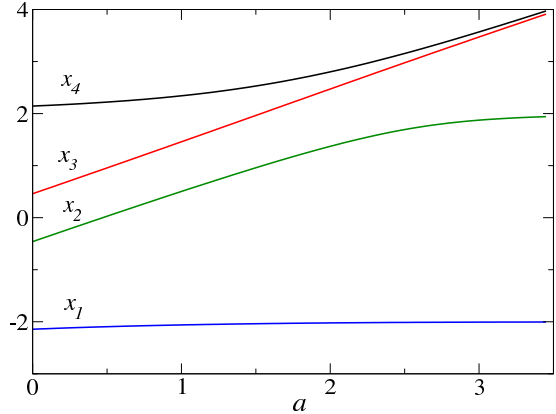}
\caption{Behavior of the edge points of the density for $m=2$ as a function of $a$ for $A=0.01$.}\label{fig:solutions}
\end{center}
\end{figure}
In Fig. \ref{fig:solutions} we plot the behavior of the endpoints $x_1,x_2,x_3,x_4$ as a function of $a$ for $A=0.01$. Note that in the limit $a\to\infty$ the pole in the potential disappears and we are left with the Gaussian potential. Indeed one can see from  Fig. \ref{fig:solutions} that for large $a\gg1$
the rightmost interval of support shrinks to zero, indicated by  $x_3\to x_4$, while the leftmost interval approaches that of the semi-circle which is located as $[-2,2]$ in our normalization. In the next subsection, we show that the lines $A=0$ and $a\to \infty$ are the only phase boundaries (i.e. there is no phase transition between two cuts and one cut at finite $a$ and non-zero $A$).

Eq. \eqref{x1} immediately confirms that the two-cut solution is inconsistent for $A\leq0$, a consequence of the unboundedness of the potential \eqref{V2def2}. Indeed, the term in brackets is always negative if $x_1<x_2<a<x_3<x_4$, and therefore the equation can never be satisfied if $A\leq 0$.

Eqs. \eqref{al2p} - \eqref{al0p} may also be used to simplify the polynomial in the numerator of $M_2(p)$ and thus the expression for the final density Eq. \eqref{rho2}:
\beq
\rho_\star^{(2)}(\la) =\frac{\la^2 -(3a-\frac12e_1)\la+\frac{1}{4} (12 a^2 -8  - 6 a  e_1 + \frac32  e_1^2 - 2 e_2)}{2\pi|\la-a|^3}\sqrt{-\la^4+e_1 \la^3-e_2 \la^2+e_3 \la -e_4}\ .\label{rhoA}
\eeq
The absolute value in the denominator reconciles different signs of the jump along the two cuts.
Together with Eqs. \eqref{x1} - \eqref{x4} this is the main result of this Section. In appendix \ref{lima0} we solve the limiting case $a\to0$ using rather Tricomi's theorem than the loop equations as an additional check.

\begin{figure}[!hbt]
\begin{center}
\includegraphics[width=0.4\columnwidth,clip=true]{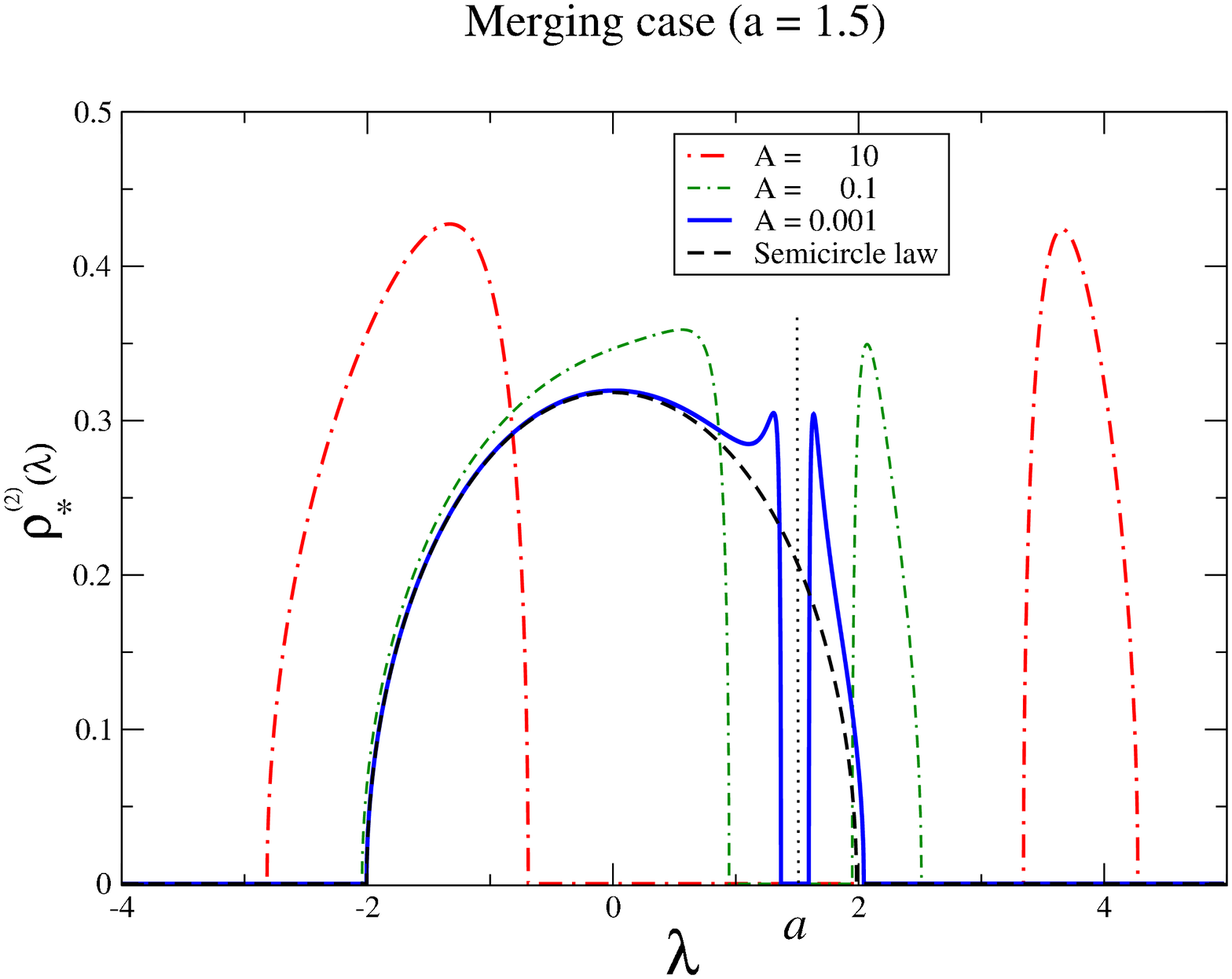}
\includegraphics[width=0.4\columnwidth,clip=true]{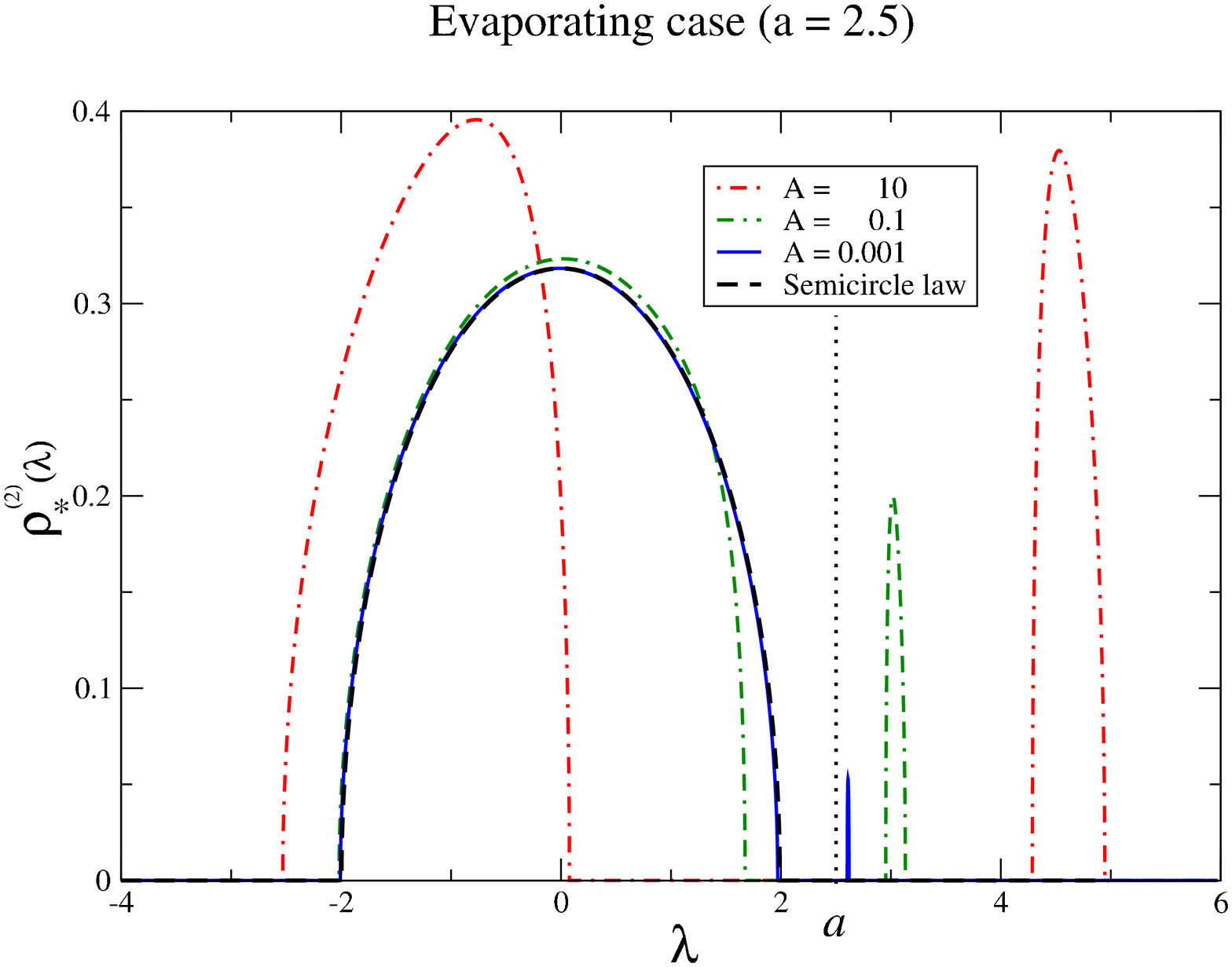}
\caption{\label{fig:merg_ev}
Graphical representation of Eq.~\eqref{rhoA} for different values of the parameter $A=10,0.1,0.001$.
Left: ``merging'' regime for  $a=1.5$; here the right cut and its density shrink to zero when $A\to0$. Right: ``evaporating'' regime, for $a=2.5$; here the two cuts merge to a single cut in the limit $A\to0$. In both cases
the semi-circle is recovered in the limit $A\to0$.}
\end{center}
\end{figure}

Fig. \ref{fig:merg_ev} illustrates the solution for the two-cut density Eq. \eqref{rhoA} for two values of $a$ and several values of $A>0$.
The boundary between the two scenarios is when the pole of the potential
is given by $a=2$, corresponding to the critical temperature $\beta=1$ in Eq. \eqref{acrit} and to the right edge of the semi-circle.

\subsection{The boundary of the two-cut phase}\label{phase}

In this Section, we discuss whether the two-cut solution \eqref{rhoA} could ever continuously evolve towards a one-cut solution as the parameters $a,A$ are varied. We know from the discussion in the preceding Section that the two-cut solution can only exist for $A> 0$ and $a\geq 0$ (finite), while for $A\to 0$ or $a\to\infty$ the solution collapses to the (one-cut) standard semicircle. As $A\to 0$, two different mechanisms exist for this limiting situation, one where the rightmost interval evaporates ($a>2$) and the other where the two intervals merge ($a<2$). Therefore, the lines $A=0$ and $a\to\infty$ constitute natural boundaries for the two-cut phase. This, however, does not in principle rule out the possibility that another phase boundary exists for $A>0,a\geq0$ (finite). We are able to show that in fact this is not the case.

Suppose that such a line does exist. This of course can only happen if the rightmost interval shrinks to zero $(x_3=x_4)$ for $a<\infty$, leaving behind an effective one-cut phase supported on\footnote{We will use the notation $y_j$ as one-cut boundaries, as opposed to $x_j$ for the two cuts. Clearly in the limiting situation considered here, we have $x_1=y_1$ and $x_2=y_2$.} $\sigma=[y_1,y_2]$, and we denote the symmetric functions of the endpoints by $f_1$ and $f_2$ below \eqref{elf}. Clearly $a>y_2>y_1$. The putative one-cut solution for the potential \eqref{V2def} will be discussed in great detail in Appendix \ref{m2example1cut}. Here we just summarize the main ingredients.
The Ansatz for the resolvent in the one-cut case which again solves a quadratic equation now reads
\beq
W_0(p)=\frac12(V'_2(p)-L_2(p)\sqrt{G(p)})\ ,
\eeq
(compare with \eqref{corrresolvent}) with the abbreviation
\beq
G(p)\ \equiv\ (p-y_1)(p-y_2)\ =\ p^2-pf_1+f_2\ .
\label{Gdef}
\eeq
It is expressed in terms of the 2 elementary symmetric functions
\beq
f_1=y_1+y_2\ \ ,\ f_2=y_1y_2\ ,\label{elf}
\eeq
while the rational function $L_2(p)$ is now given by
\beq
L_2(p)=\frac{\ga_3p^3+\ga_2p^2+\ga_1p+\ga_0}{(p-a)^3}
\ ,
\label{L2P2main}
\eeq
The corresponding putative one-cut density (indicated by the superscript $^{(1)}$) reads
\beq
\rho_\star^{(1)}(\la) =\ \frac{1}{2\pi} L_2(\la)
\sqrt{(\la-y_1)(y_2-\la)}\ ,\ \ \la\in[y_1,y_2]\ .
\eeq

Now it is possible to determine that density by putting $x_3=x_4$ into the two-cut density and pertinent equations. However, this leads to a contradiction in the resulting one-cut setting (unless $a\to\infty$) for $A>0$. Therefore such phase boundary between the two-cut and the one-cut phase does not exist for $a<\infty$. Indeed for $x_3=x_4$ the following holds
\begin{align}
e_1 &=f_1+2x_3\ ,\label{e1f}\\
e_2 &=f_2+2 x_3 f_1+x_3^2\ ,\label{e2f}\\
F(a) &= G(a) (a-x_3)^2\ .\label{FasG}
\end{align}
Replacing these expressions into \eqref{x1}, \eqref{x2} and \eqref{x3}, after lengthy algebra and many simplifications we precisely arrive at equations
\eqref{y120def} and \eqref{y12def} that need to be satisfied by the endpoints $y_1$ and $y_2$ of the one-cut density, supplemented by the extra condition
\beq
t\left(\frac{a^2 f_1}{2}-\frac{3 a f_1^2}{4}+a f_2+4 a+\frac{5 f_1^3}{16}-\frac{3
   f_1 f_2}{4}-f_1\right)=-\frac{a f_1}{2}+\frac{3 f_1^2}{8}-\frac{f_1}{2 t}-\frac{f_2}{2}+\frac{1}{t^2}-2\ ,\label{firstgood}
\eeq
with
\beq
a-x_3=\frac{1}{t}\ .
\eeq
At the same time, the density $\rho^{(2)}(\lambda)$ converges to the density $\rho^{(1)}(\lambda)$ upon changing $e_j$ into $f_j$. Therefore when the rightmost interval shrinks $(x_3=x_4)$ the two-cut solution precisely transforms into the putative one-cut solution (see Appendix \ref{m2example1cut}) where the endpoints satisfy the two equations \eqref{y120def} and \eqref{y12def} as expected. However, there is an extra condition \eqref{firstgood} that needs to hold, which yields a relation between $x_3$ (the collapse point) and $a$. It can be shown using Mathematica that this relation violates the ordering constraint $x_3>a>y_2>y_1$ and $A>0$, implying that the transition between two-cut and one-cut phase does not take place anywhere else than for $A=0$ or $a\to\infty$.

\setcounter{equation}{0}
\section{Numerical simulations}\label{sec:numerical}
In order to verify numerically the solution \eqref{rhoA} for the
spectral density, one can implement Monte Carlo simulations
exploiting the analogy between the eigenvalues of the random matrix
ensemble and particles interacting with a two-dimensional Coulomb
potential that are constrained to move on a line. More specifically, the
system of particles $\{\lambda \}$ evolves according to the
following Hamiltonian
\begin{equation}
E(\{\lambda\})=\frac{1}{2}\sum_{i}\lambda^{2}_{i}-\sum_{i,j}\log\vert\lambda_{i}-\lambda_{j}\vert+\sum_i\frac{2A}{(\lambda_{i}-a)^{2}} \ ,\label{montecarlo_energy}
\end{equation}
under the additional constraint (zero trace condition)
\begin{equation}
\sum_{i}\lambda_{i}=0\ .\label{zero_trace_constraint}
\end{equation}
At each step a pair of particles $(\lambda_{i},\lambda_{j})$
is chosen at random and a change in their position
$(\lambda_{i}+\Delta \lambda,\lambda_{j}-\Delta \lambda)$ is proposed,
where $\Delta \lambda$ is a Gaussian random variable, with
zero mean and variance $\epsilon$. With this choice, if the constraint
\eqref{zero_trace_constraint} is satisfied for the initial condition,
it keeps holding throughout the dynamics. The suggested change in
the particles position corresponds to a change in the energy
$\Delta E$ and is accepted with probability $\textrm{min}(e^{-\beta_D
  \Delta E},1)$, as the standard rule for the Metropolis algorithm. By tuning the parameter $\epsilon$ one can optimize
the convergence rate of the algorithm. Generally this parameter is fixed in
such a way that the rejection rate is approximately equal to $1/2$.\\

\begin{figure}[h!bt]
\begin{center}
\includegraphics[width=0.4\columnwidth,clip=true]{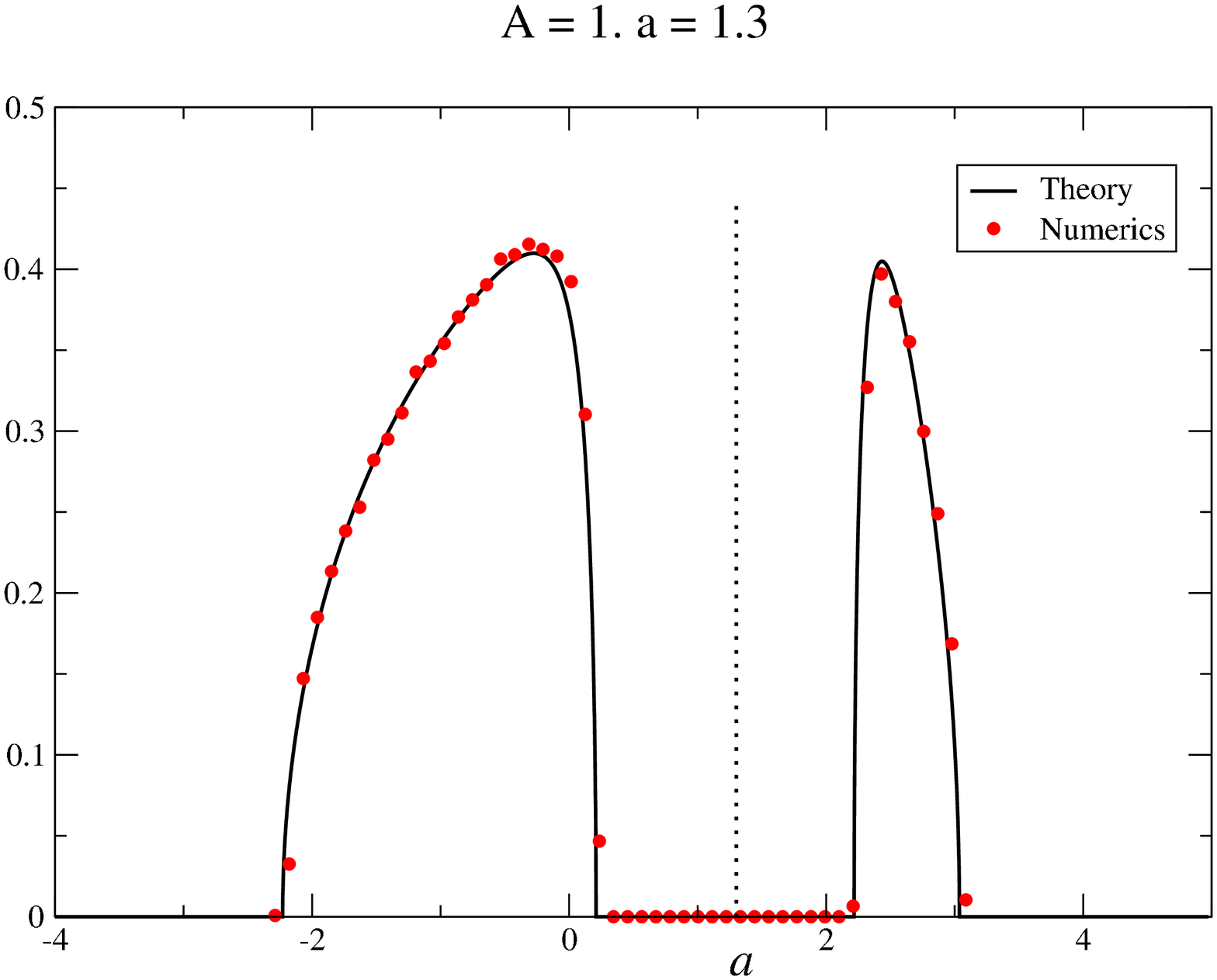}
\includegraphics[width=0.4\columnwidth,clip=true]{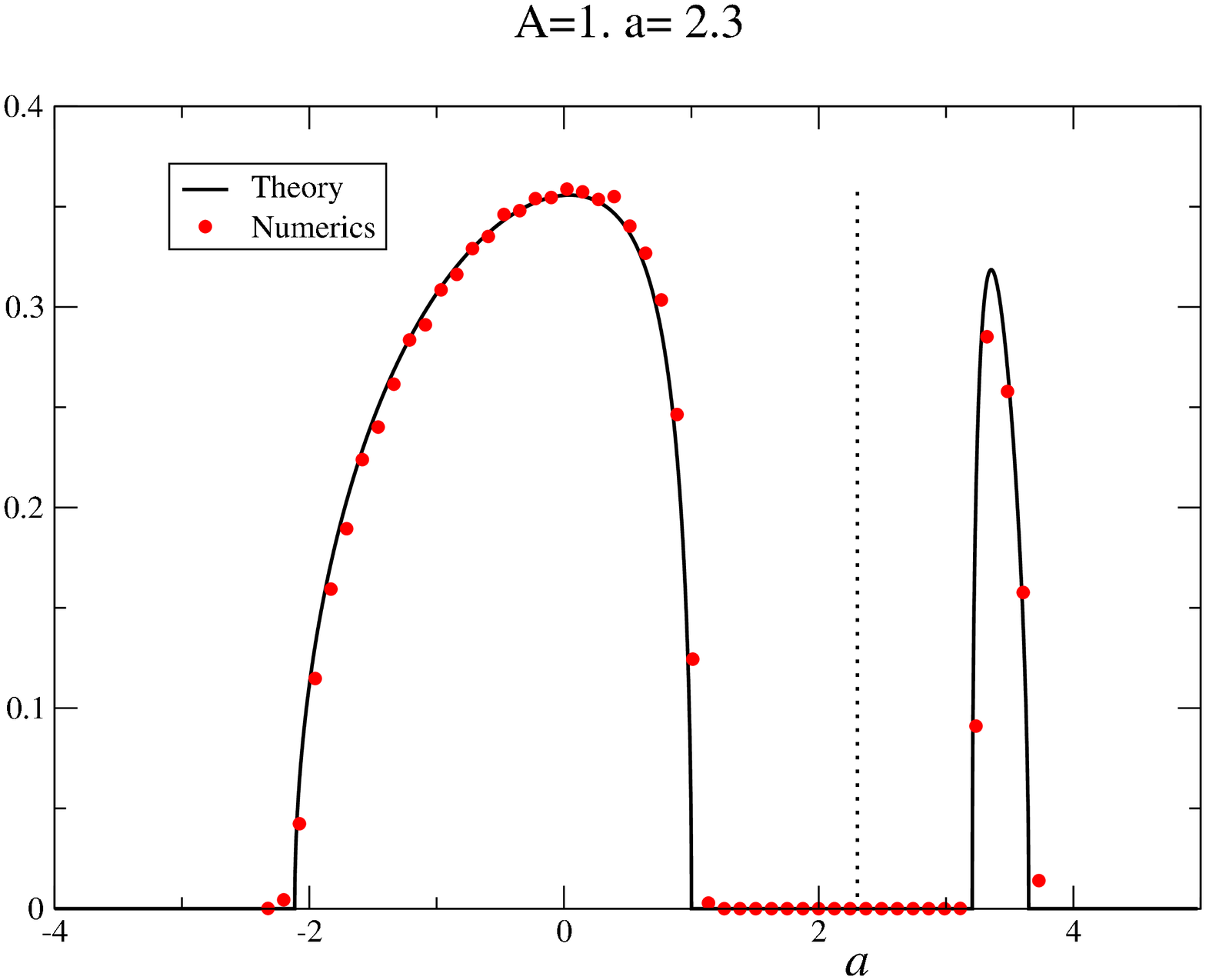}
\caption{\label{fig:check} Numerical verification of Eq.~\eqref{rhoA} for two different values of the parameters, with $N=50$.}
\end{center}
\end{figure}

The presence of the singularity in the confining potential leads to 
two different supporting intervals for the density, and the Coulomb
interaction in Eq.~\eqref{montecarlo_energy} makes these intervals well
separated for any finite value of $A$, (see, for instance
Fig.~\ref{fig:merg_ev}). Hence, the probability of observing transitions from one support to
the other is exponentially small and the choice of the initial condition
plays a relevant role for the convergence time of the algorithm.
For these reasons, to properly test the analytical results we
have applied two different recipes:
\begin{itemize}
\item We have numerically computed the conditional average
  $\left<E(\{\lambda\})\vert N_{l}\right>$ as a function of different
  values of $N_{l}$ for fixed values of $A$ and $a$, $N_{l}$ being the
  number of particles to the left of $a$ (in other
  words, the number of particles in the left interval). The
  lowest value of the function $\left<E(\{\lambda\})\vert N_{l}\right>$
  gives an estimate of the $N^{*}_{l}$ that must be chosen as
  initial condition in order to ensure the fastest convergence of the
  algorithm towards the equilibrium distribution.

\item We have used an appropriate ``annealing'' procedure, putting
  a cut-off in the energy differences. More specifically, starting
  from a random 
configuration that satisfies zero trace we have considered a thermalization
  procedure of duration $T_{term}$,
  where the Monte Carlo has been performed according to the following step-dependent rule
\begin{equation}
\Delta E_{ef\!f}=\left\{ \begin{array}{cr}\Delta E & \textrm{if } \Delta E < E_{max}(t)\\ E_{max}(t) & \textrm{if } \Delta E > E_{max}(t)\end{array}\right.
\end{equation}
We have taken $E_{max}(t)=\kappa_{0}\frac{t}{T_{term}}$, where
$\kappa_{0}$ is a parameter in the interval $[1,10]$. These cycles of $T_{term}$
can be repeated $n$ times in order to find configurations with low energy. Such configurations are used as a starting point for the
equilibrium Metropolis algorithm with the ordinary energy difference
$\Delta E$. The purpose of the cutoff $\Delta E_{ef\!f}(t)$ is precisely to artificially lower the energy barrier for short times $t\ll T_{term}$ and
to allow jumps of particles from one interval to the other that would otherwise be extremely rare.

\end{itemize}
Following both procedures, the algorithm converges to the proper equilibrium configuration and all analytical predictions are well confirmed by numerical simulations (see Fig.~\ref{fig:check}).

\section{Conclusions}\label{sec:conclusions}

In summary, we have computed the large $N$ spectral density $\rho_\star(\lambda)$ (see \eqref{rhoinitial}) for an invariant ensemble of $N\times N$ random matrices, where the standard Gaussian weight is distorted by an extra pole of order $m$ and the ensemble is traceless on average. This density generally consists of two sets of eigenvalues lying on either side of the pole, and is obtained solving the loop equation with the additional constraint of vanishing trace on average. We proved that no single-cut phase exists anywhere in the $(A,a)$ plane except for $A=0$ or $a\to\infty$, where $A$ tunes the strength of the additional singular interaction in the potential and $a$ is the location of the pole. This study (for the orthogonal case and $m=2$) is motivated by an application to the physics of the Sherrington-Kirkpatrick mean-field model of spin glasses. Deep in the paramagnetic phase, the spin glass susceptibility, a standard indicator of the onset of a glassy phase, depends on the eigenvalues of the inverse susceptibility matrix, which is nothing but the Hessian of the magnetization-dependent free energy at the relevant minimum. In the TAP approach, the free energy is given by a finite sum of terms (see \eqref{3b}) and the inverse susceptibility matrix at the paramagnetic minimum ($m_i=0\quad\forall i$) acquires a particularly simple form \eqref{eq:hes} in terms of the elements of the coupling matrix $\{x\}$. This way, under very mild assumptions the inverse susceptibility matrix can just be written as a linear statistics on the eigenvalues of $\{x\}$. Drawing these matrix elements $\{x\}$ at random from the GOE ensemble makes it possible to apply standard Coulomb fluid techniques in studying the distribution of the spin glass susceptibility for large system sizes. The free energy of the associated fluid is precisely determined (to leading order in $N$) by the partition function of our RMT (see \eqref{SP}), where $a$ is related in a simple way to the inverse temperature of the SK model \eqref{acrit}. The average spectral density of our RMT is therefore a crucial ingredient to determine the rate (or large deviation) function for the probability of rare fluctuations of this susceptibility in the paramagnetic phase. The analytical prediction for the average density has been accurately verified with sophisticated Monte Carlo simulations. 

The present work provides evidence that standard RMT techniques such as the Coulomb fluid analogy and loop equations might be of great usefulness in a spin glass setting. Moreover, a few clear directions of research naturally emerge from this study: first, it will be crucial to determine the rate function explicitly by inserting the average spectral density \eqref{density} into the action \eqref{actionS} and by solving the corresponding integrals if possible. Next, this analytical rate function should be compared with $\iota)$ high-precision numerical simulation of the distribution of the square of the overlap in the SK model, possibly recording rare events where such random variable is much larger than its typical value, and $\iota\iota)$ accurate numerical simulations of the distribution of $\chi_{\mathrm{SG}}^x(\beta,N)$ as defined in \eqref{chilinear} by sampling large GOE matrices. This will also constitute a check of the validity of the TAP equations in the paramagnetic regime (obtained by neglecting higher order in $\alpha$ and used to define the inverse susceptibility matrix \eqref{eq:hes}), as well as the (very mild) assumption of neglecting correlations between the diagonal and off diagonal terms in the coupling matrix (see \cite{zarinelli2}).

\begin{acknowledgments}
We are indebted to Michele Castellana and Pierfrancesco Urbani for very helpful discussions on the spin-glass physics and for continuous advice and support. We are grateful to Aurelien Decelle, Gino Del Ferraro, Silvio Franz, Mario Kieburg, Luca Leuzzi, Giacomo Livan, Cristophe Texier and Elia Zarinelli for illuminating discussions at various stages of this project. D.V. acknowledges support of the LPTMS postdoc fundings during the early stage of this project. We acknowledge partial support from Labex/PALM (project RANDMAT) (P.V.)
and from the SFB $|$ TR12 ``Symmetries and Universality
in Mesoscopic Systems'' of the German research council DFG (G.A.).
\end{acknowledgments}
\appendix


\setcounter{equation}{0}\section{Symmetric limit $a\to0$ from Tricomi's theorem}\label{lima0}

In this Appendix, we compute the two-cut spectral density in the limit $a\to 0$ using an alternative method. This constitutes an independent check of previous results, and again confirms that no stable two-cut solution exists when $A\leq 0$.

In the case $a\to 0$, the potential Eq. \eqref{V2def2} and therefore the density are even functions. The support can then be written as $\sigma=[-x_4,-x_3]\cup [x_3,x_4]$. The density is the solution of the following singular integral equation of Tricomi type
Eq. \eqref{singint} (with $C=0$)\footnote{Because the limiting density is symmetric the zero-trace constraint is automatically satisfied.}
\begin{equation}
\mathrm{Pr}\int_\sigma\frac{\rho_\star(x^\prime)}{x-x^\prime}dx^\prime = \frac12 x-\frac{2A}{x^3}\ ,\quad{x\in\sigma}
\end{equation}
or, explicitly
\begin{equation}
\int_{-x_4}^{-x_3}\frac{\rho_\star(x^\prime)}{x-x^\prime}dx^\prime + \mathrm{Pr}\int_{x_3}^{x_4}\frac{\rho_\star(x^\prime)}{x-x^\prime}dx^\prime = \frac12 x-\frac{2A}{x^3},\qquad x\in (x_3,x_4)\ .
\end{equation}
Changing variables $x^\prime\to -x^\prime$ in the first integral and using parity, $\rho_\star(x)=\rho_\star(-x)$, we get
\begin{equation}
\int_{x_3}^{x_4}\frac{\rho_\star(x^\prime)}{x+x^\prime}dx^\prime + \mathrm{Pr}\int_{x_3}^{x_4}\frac{\rho_\star(x^\prime)}{x-x^\prime}dx^\prime = \frac12 x-\frac{2A}{x^3},\qquad x\in (x_3,x_4)\ ,
\end{equation}
or
\begin{equation}
2x\ \mathrm{Pr}\int_{x_3}^{x_4}dx^\prime \frac{\rho_\star(x^\prime)}{x^2-x^{\prime 2}}= \frac12 x-\frac{2A}{x^3},\qquad x\in (x_3,x_4)\ .
\end{equation}
Denoting $x^2=y$ and $x^{\prime 2}=y^\prime$ we get
\begin{equation}
\mathrm{Pr}\int_{x_3^2}^{x_4^2}dy^\prime \frac{\phi(y^\prime)}{y-y^\prime}=\frac12-\frac{2A}{y^2}\label{refor}\ ,
\end{equation}
where $\phi(x)=\rho_\star(\sqrt{x})/\sqrt{x}$. The reformulation \eqref{refor} makes it possible to use the single-support inversion formula \cite{tricomi}
\begin{equation}
\phi(x)=\frac{1}{\pi\sqrt{(x-x_3^2)(x_4^2-x)}}\left[h-\ \mathrm{Pr}\int_{x_3^2}^{x_4^2}\frac{dt}{\pi}\frac{\sqrt{(t-x_3^2)(x_4^2-t)}}{x-t}\left(\frac12-\frac{2A}{t^2}\right)\right]\ ,
\end{equation}
where $h$ is an arbitrary constant. Evaluating the principal value integral, we get
\begin{equation}
\phi(x)=\frac{1}{\pi\sqrt{(x-x_3^2)(x_4^2-x)}}\left[h-\frac12
   \left(x-\frac{x_3^2}{2}-\frac{x_4^2}{2
   }\right)+\frac{A \left(x_4^2 \left(x-2
   x_3^2\right)+x x_3^2\right)}{x^2
   x_3 x_4}\right]\ .
\end{equation}
Imposing the normalisation $\int_{x_3}^{x_4}dx^\prime \rho_\star(x^\prime)=1/2$ (which is equivalent to $\int_{x_3^2}^{x_4^2}dx^\prime \phi(x^\prime)=1$) yields $h=1$, and the requirement $\rho_\star(x_3)=\rho_\star(x_4)=0$ yields the two conditions
\begin{equation}
-\frac{A x_4}{x_3^3}+\frac{A}{x_3
   x_4}-\frac{ x_3^2}{4}+\frac{
   x_4^2}{4}+1=0\ ,
\label{constra}
\end{equation}
as well as the equation obtained by swapping $x_3\leftrightarrow x_4$. The final expression for the density then reads
\begin{equation}
\rho_\star (\lambda) =
\frac{2 \left(\lambda ^2
   \left(x_3^2+x_4^2\right)+2 x_3^2
   x_4^2\right)}{\pi  |\lambda| ^3
   \left(x_3^2-x_4^2\right)^2}
\sqrt{(x_3^2-\lambda^2 )(\lambda^2 -x_4^2)}\ .
\label{rhofinalb}
\end{equation}
A similar calculation can be done for any even $m$. 

In order to compare to the solution from loop equations Eq. \eqref{rhoA} for $m=2$ in the limit $a\to0$ let us express the conditions \eqref{constra} in terms of the symmetric functions $e_i$.
Taking $a\to 0$ we have $x_1=-x_4$ and $x_2=-x_3$, and $0<x_3<x_4$ in order to have two cuts\footnote{Note that because we are now dealing with a symmetric potential the corresponding two-cut solution is no longer an underdetermined system due to symmetry.}.
From \eqref{elem} we thus obtain
\beql
0&=&e_1=e_3\ ,\\
e_2&=&-x_3^2-x_4^2<0\ ,\\
e_4&=&x_3^2 x_4^2\geq0\ .
\label{esym}
\eeql
Taking sum and difference of Eq. \eqref{constra} and its counterpart
with exchanged $x_3\leftrightarrow x_4$ we have respectively
\beql
0&=& -A(e_2^2-4e_4)+2e_4^{3/2}\\
0&=& (x_4^2-x_3^2)\left(2Ae_2+e_4^{3/2}\right)\ .
\eeql
Because the first factor in the second equation cannot vanish - else both cuts would be zero - we conclude
\beq
0=2Ae_2+e_4^{3/2}\ .
\label{condX}
\eeq
This can be used in the first equation of \eqref{esym}, and if $A\neq0$ we arrive at
\beq
e_4=e_2+\frac14 e_2^2\ .
\label{e42}
\eeq
If we now compare the expression for the density \eqref{rhoA} derived from loop equations in the limit of $a\to0$,
\begin{align}
\rho_\star(\la) &=\frac{1}{2\pi} \frac{\la^2 +\frac{1}{4} (-8 - 2 e_2)}{|\la|^3}\sqrt{-\la^4-e_2 \la^2-e_4}\ ,
\label{rhoAa}
\end{align}
we find a perfect matching with Eq. \eqref{rhofinalb} after some algebra.

The condition Eq. \eqref{condX} also allows to see that the two-cut solution is inconsistent for $A<0$ in this symmetric limit. Indeed, given the signs of  $e_2$ and $e_4$, Eq. \eqref{condX} can only have real solutions for the endpoints $x_3$ and $x_4$ for $A>0$. This confirms our analysis for general $a\geq0$.

As a last check we can recover the conditions for the edges of the semicircle. From \eqref{condX} setting $A=0$ is equivalent to $x_3=0$, that is the merging of the two cuts (we have $0<x_4$ for a finite support). It then follows from \eqref{e42} that $x_4=2(=-x_1)$, as we need for the semicircle.


\setcounter{equation}{0}\section{The one-cut solution and its incompatibility with zero trace}\label{m2example1cut}

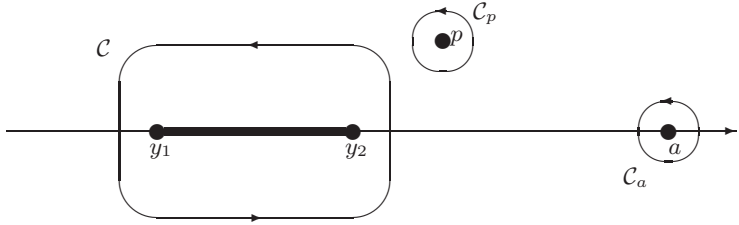
\begin{figure}[h]
\unitlength1cm
\begin{picture}(12.2,5.5)
\linethickness{1.0mm}
\put(2.1,2.3){\line(1,0){2.4}}
\thinlines
\put(0,2.3){\line(1,0){1.9}}
\put(4.7,2.3){\line(1,0){1}}
\put(6.9,2.3){\line(1,0){1}}
\put(7.7,2.3){\vector(1,0){2}}
\multiput(2,2.3)(2.6,0){2}{\circle*{0.2}}
\put(5.8,3.5){\circle*{0.2}}
\put(5.9,3.5){$p$}
\multiput(5.8,3.5)(3,-1.2){2}{\oval(0.8,0.8)[l]}
\multiput(5.8,3.5)(3,-1.2){2}{\oval(0.8,0.8)[r]}
\multiput(5.8,3.9)(3,-1.2){2}{\vector(-1,0){0.1}}
\put(5.7,2.3){\line(1,0){2}}
\put(8.8,2.3){\circle*{0.2}}
\put(8.8,2){$a$}
\put(8.2,1.6){${\cal C}_{a}$}
\put(6.2,3.8){${\cal C}_{p}$}
\put(1.9,2){$y_{1}$}
\put(4.5,2){$y_{2}$}
\put(2.,2.3){\oval(1.0,2.3)[l]}
\put(4.6,2.3){\oval(1.0,2.3)[r]}
\put(2.,3.45){\line(1,0){1.2}}
\put(3.4,1.15){\line(1,0){1.2}}
\put(4.6,3.45){\vector(-1,0){1.4}}
\put(2.,1.15){\vector(1,0){1.4}}
\put(1.2,3.3){${\cal C}$}
\end{picture}
\caption{\label{contour1cut}The contour of integration
         $ {\cal C}$ for one cut, the location of the pole of the
  potential at $z=a$ and the argument of the resolvent at $z=p$.
}
\end{figure}
\begin{figure}[h!bt]
\begin{center}
\includegraphics[width=0.7\columnwidth,clip=true]{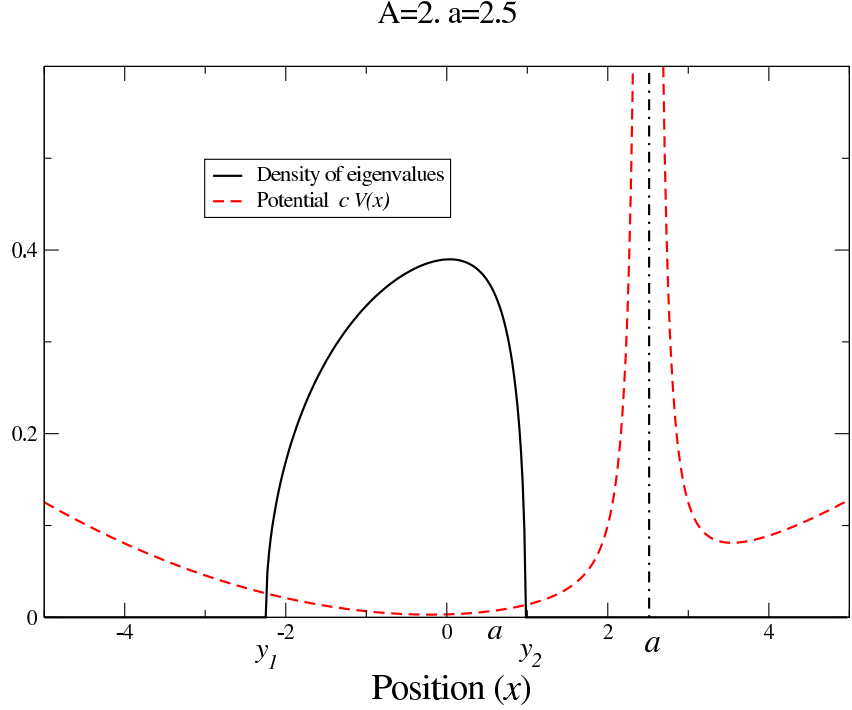}
\caption{Non-traceless one-cut solution for the density \eqref{1cutd} with $A=2$ and $a=2.5$, together with the confining potential $V_2(x)$ rescaled by $c=0.1$ as in Fig. \ref{fig:density1}. While this is only an acceptable solution without imposing the zero-trace constraint, we have seen also numerically that after imposing the constraint both minima are filled, which is the two-cut solution.}\label{fig:onecut}
\end{center}
\end{figure}

In this Appendix we repeat the calculation from subsection \ref{m2example} with $m=2$, but this time assuming a single interval support $\sigma=[y_1,y_2]$. Using again the loop equation machinery, we are led to two equations (see \eqref{y120def} and \eqref{y12def} below) that fix the endpoints $y_1,y_2$ of the support as functions of $a,A$, while the general expression for the one-cut solution is given in \eqref{rho21} (using \eqref{L2P2} and subsequent equations for $\gamma_j$). Such expressions yield a density $\iota)$ where the traceless constraint has not yet been imposed, and $\iota\iota)$ 
which is not guaranteed to be positive definite (this depends on the specific choice of the parameters $a,A$), since the density might develop a further zero inside or at the edge of the support $\sigma$. Once the traceless constraint is imposed, however, a further equation \eqref{traceless} relating $y_1,y_2$ is found, that singles out one line $A=A(a)$ in the $(A,a)$ plane where the putative (one-cut and traceless) solution must live. However, it turns out that on such line the one-cut density is \emph{never} positive definite (unless $A=0$). Therefore, an acceptable traceless one-cut density does not exist anywhere in the $(a,A)$ plane (unless $A=0$ or $a\to\infty$), as already anticipated in Section \ref{phase}. In view of this negative result for the physically relevant traceless case, we refrain from giving more unnecessary details on the positivity of the non-traceless density (i.e. without imposing the further condition \eqref{traceless}). We will however include a picture below (Fig. \ref{fig:onecut}) for a specific choice of $a,A$ where this (non-traceless) density \emph{formally}\footnote{For such acceptable values of $a,A$ we would have formal coexistence of the (traceless) two-cut and (non-traceless) one-cut phases, the true phase being selected by free energy minimization. However, we will not dwell on this non-traceless case in the following.} does exist.

\begin{figure}[h!bt]
\begin{center}
\includegraphics[width=0.7\columnwidth,clip=true]{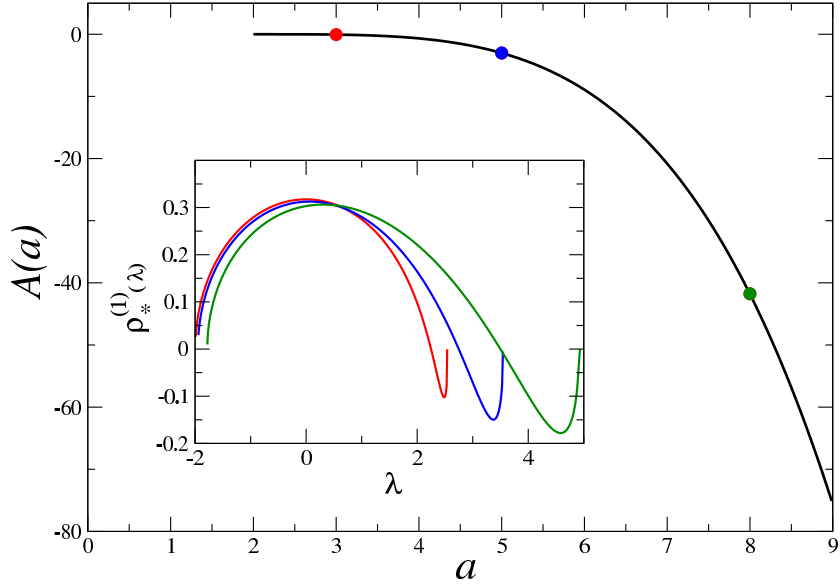}
\caption{Plot of the critical line $A=A(a)$ where the traceless one-cut solution must live. Inset: on this line, the putative density is never positive definite. For $a=2$, $A(2)=0$ and the density recovers the semicircle. There is no solution for the critical line for $a<2$.}\label{fig:critical}
\end{center}
\end{figure}

\begin{figure}[h!bt]
\begin{center}
\includegraphics[width=0.7\columnwidth,clip=true]{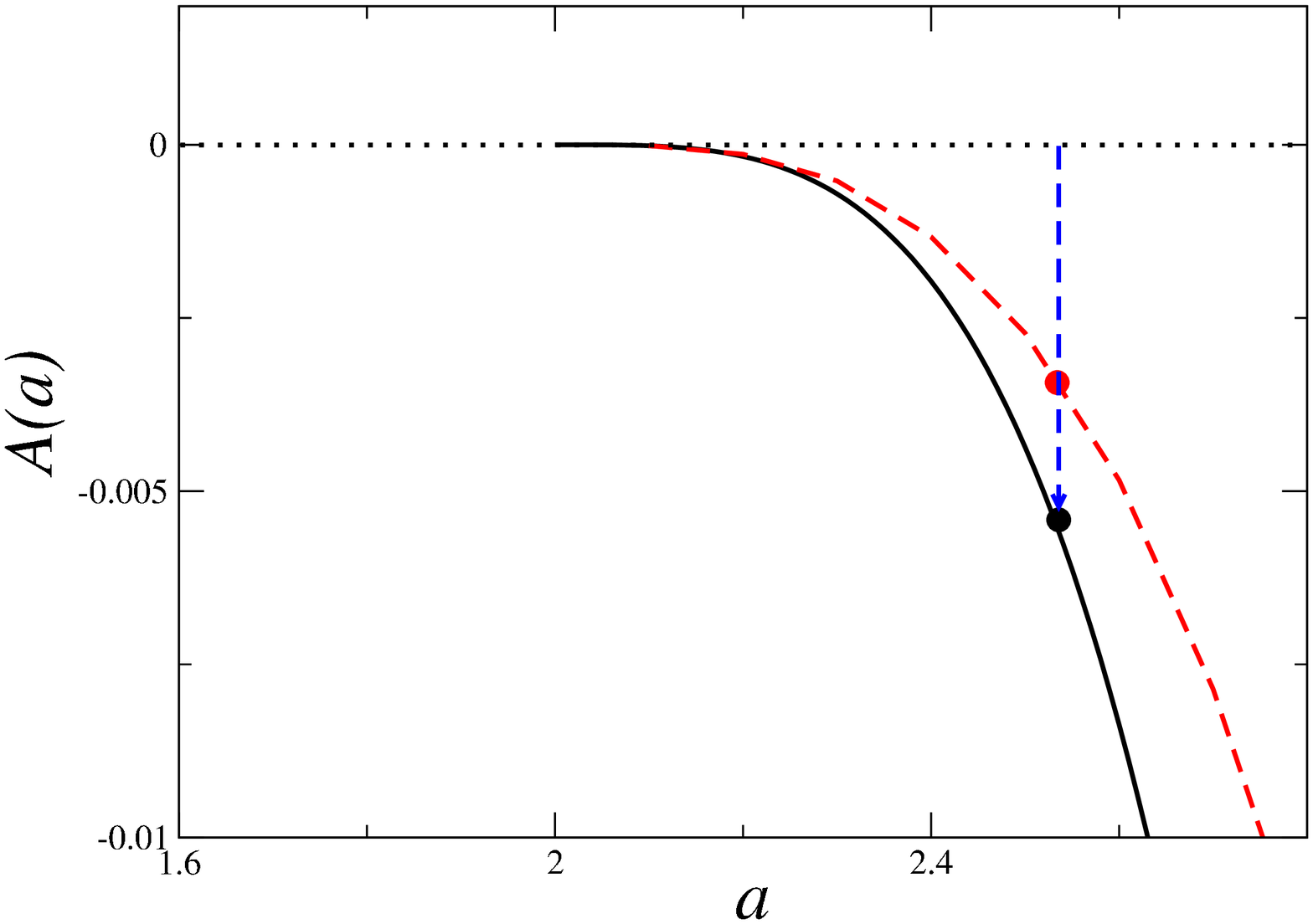}
\caption{Plot of the critical line $A=A(a)$ where the traceless one-cut solution must live (black solid) together with the line on which $L_2(y_2)=0$ (dashed red). At fixed $a>2$, coming from $A=0$ and following the dashed blue arrow, one meets \emph{first} the line at which a zero of the density develops at the right edge, and only \emph{later} the critical line of existence of a traceless one-cut phase, where the zero moves to the left, inside the support. This convincingly corroborates the non-existence of a positive definite and traceless one-cut density.}\label{fig:twocriticallines}
\end{center}
\end{figure}

We start by recalling some notation that was already used in Section \ref{phase}.
The potential is again given by
\beq
V_2(x)=\frac12 x^2+\frac{2A}{(x-a)^2}\ .
\eeq
The one-cut Ansatz is parametrized by $\sigma=[y_1,y_2]$, and we denote the symmetric functions of the endpoint by $f_1$ and $f_2$ below.
For stability reasons we will also require that $a>y_2>y_1$.

Since the computation is very similar to the one presented in subsection \ref{planar} we will be brief. The integration contour in the loop equation \eqref{loop} has to be replaced by the contour given in Fig. \ref{contour1cut}.
The Ansatz for the resolvent which again solves a quadratic equation now reads
\beq
W_0(p)=\frac12(V'_2(p)-L_2(p)\sqrt{G(p)})\ ,
\label{ansatz1cut}
\eeq
with the abbreviation
\beq
G(p)\ \equiv\ (p-y_1)(p-y_2)\ =\ p^2-pf_1+f_2\ .
\label{fdef}
\eeq
It is expressed in terms of the 2 elementary symmetric functions
\beq
f_1=y_1+y_2\ \ ,\ f_2=y_1y_2\ .
\eeq
The corresponding one-cut density reads
\beq
\rho_\star^{(1)}(\la)\ =\ \frac{1}{2\pi} L_2(\la)
\sqrt{(\la-y_1)(y_2-\la)}\ ,\ \ \la\in[y_1,y_2]\ .
\label{rho21}
\eeq
Eq. \eqref{Mcontour2} with the one-cut contour depicted in Fig. \ref{contour1cut} allows us to determine the coefficients of the rational function $L_2(\la)$:
\beql
L_2(p)&=&\left.\left(\frac{2A}{(w-p)\sqrt{(w-y_1)(w-y_2)}}
\right)^{\prime\prime}\right|_{w=a}+1\nn\\
&=&\frac{4A}{(a-p)^3G(a)^{1/2}}+\frac{2AG'(a)}{(a-p)^2G(a)^{3/2}}
-\frac{AG''(a)}{(a-p)G(a)^{3/2}}+\frac{3AG'(a)^2}{2(a-p)G(a)^{5/2}}\ +1\ .
\label{L2}
\eeql
Its form agrees with the corresponding 2-cut expression Eq. \eqref{M2}, apart from the last term coming from contribution at infinity which is now non-zero.
Note that in Eq. \eqref{ansatz1cut} we have chosen the branch of the square root $\sqrt{G(w)}\sim w$ for $|w|\gg1$. Because we only have one cut and $a>y_2$ one has that $\sqrt{G(a)}>0$ is still positive. Hence there is no need here to explicitly display the sign of the branch (in contrast to two cuts), and we can write $\sqrt{G(a)}=G(a)^{1/2}$ both being the principal branch.

The coefficients in the numerator of $L_2(p)$ are given by
\beq
L_2(p)=\frac{\ga_3p^3+\ga_2p^2+\ga_1p+\ga_0}{(p-a)^3}
\ ,
\label{L2P2}
\eeq
and can be simply read off comparing Eq. \eqref{L2} and \eqref{L2P2}: 
\beql
\ga_3&=& 1\ ,\label{ga3}\\
\ga_2&=&-3a+\frac{A}{G(a)^{5/2}}\left(-\frac32 G'(a)^2+G''(a)G(a)\right)\nn\\
&=& -3a + \frac{A}{2G(a)^{5/2}}(-8 a^2+8 a f_1-3 f_1^2+4 f_2)\ ,
\label{ga2}\\
\ga_1&=&3a^2+\frac{A}{G(a)^{5/2}}(2G(a)G'(a)-2aG(a)G''(a)+3aG'(a)^2)
\nn\\
&=&3a^2+\frac{A}{G(a)^{5/2}}(12 a^3-14 a^2 f_1+5 a f_1^2-2 f_1 f_2)\ ,
\label{ga1}\\
\ga_0&=&-a^3+\frac{-A}{G(a)^{5/2}}
\left(4G(a)^2+2aG(a)G'(a)-a^2G(a)G''(a)+\frac32a^2G'(a)^2\right)\nn\\
&=& -a^3+\frac{-A}{G(a)^{5/2}}(12 a^4-18 a^3 f_1 +\frac{15}{2}a^2 f_1^2+10 a^2 f_2-10 a f_1 f_2+4 f_2^2) \label{ga0}
\ .
\eeql
The positions of the two endpoints $y_{1},y_2$ as functions of $a$ and $A$ are again determined by the asymptotic expansion \eqref{W-pasympt} of the planar resolvent, Eq. \eqref{ansatz1cut}. We will express this expansion in terms of the coefficients $\ga_j$ we have just determined.
\beql
{\cal O}(p):\ \ 0&=& 1-\ga_3\nn\\
&\Leftrightarrow& \ga_3=1 \ ,\qquad
\label{Op-1c}\\
{\cal O}(1):\ \ 0&=& \ga_3(3a-f_1/2)+\ga_2 \nn\\
&\Leftrightarrow&\ga_2=\frac12f_1-3a \qquad
\label{O1-1c}\\
{\cal O}(p^{-1}):\ \ 1&=& -\frac{1}{16}\left(8\ga_1+\ga_2(24a-4f_1)
+\ga_3(48a^2-12af_1-f_1^2+4f_2)\right)\ ,
\nn\\
&\Leftrightarrow&\ga_1=-2+\frac38f_1^2-\frac32af_1+3a^2-\frac12f_2
\ .
\label{asymp-cond1cut}
\eeql
The first equation is identically satisfied. This leaves us with two equations for the two endpoints which we give again explicitly,
\beql
f_1
&=& 
\frac{A}{G(a)^{5/2}}(-8 a^2+8 a f_1-3 f_1^2+4 f_2)\label{y120def}\\
-2+\frac38f_1^2-\frac32af_1
-\frac12f_2\ ,
&=&
\frac{A}{G(a)^{5/2}}(12 a^3-14 a^2 f_1+5 a f_1^2-2 f_1 f_2)\ .
\label{y12def}
\eeql
Together with Eq. \eqref{rho21} which we have again simplified inserting the expressions for some of the $\gamma_j$, this leads to the density 
\beq
\rho_\star^{(1)}(\la)\ =\ \frac{\la^3+\la^2\left(\frac12f_1-3a\right)
+\la\left(-2+\frac38f_1^2-\frac32af_1+3a^2-\frac12f_2\right)+\ga_0}{2\pi(\la-a)^3}
\sqrt{-\la^2+f_1\la-f_2}
\ ,\ \ \la\in[y_1,y_2]\ .\label{1cutd}
\eeq
This is the solution for the one-cut Ansatz, so far without imposing neither the zero trace constraint, nor the condition of positivity of the density.

In principle we could now impose the positivity constraint (i.e. that no further zero develops inside or at the endpoints of the support $\sigma$) and determine the phase boundaries of this one-cut solution, in analogy to subsection \ref{phase}. However, we will not follow this route now and rather first impose the (physically relevant) zero trace constraint.
Following Eq. \eqref{W-pasympt} for the asymptotic expansion of $W_0(p)$ in $p$ we obtain a second equation for $\gamma_0$ from this constraint:
\beql
&&{\cal O}(p^{-2}):\nn\\ 
0&=& \ga_0+3a\ga_1+6a^2\ga_2+10a^3\ga_3-\frac12(\ga_1+3a\ga_2+6a^2\ga_3)f_1
+(\ga_2+3a\ga_3)\left(-\frac18f_1^2+\frac12 f_2\right)+\ga_3\left(-\frac{f_1^3}{16}+\frac{f_1f_2}{4}\right) \nn\\
&\Leftrightarrow&\ga_0=6a-f_1+\frac32af_2-a^3+\frac32a^2f_1-\frac98 a f_1^2 -\frac34 f_1f_2+\frac{5}{16}f_1^3\ ,
\label{ga0phase}
\eeql
which together with Eq. \eqref{ga0} reads
\beql
&&\frac{-A}{G(a)^{5/2}}
\left(4G(a)^2+2aG(a)G'(a)-a^2G(a)G''(a)+\frac32a^2G'(a)^2\right)\nn\\
&=&6a-f_1+\frac32af_2+\frac32a^2f_1-\frac98 a f_1^2 -\frac34 f_1f_2+\frac{5}{16}f_1^3\ .\label{traceless}
\eeql
We now have three equations for the two endpoints and thus the system is overdetermined. This should project onto a line in the allowed phase space of the one-cut solution {\it without} zero trace  constraint. This line $A=A(a)$ where the traceless one-cut solution must live is plotted in Fig. \ref{fig:critical}. In the inset, the obtained density is shown to be unacceptable, as it is never positive definite, unless for $a=2$ where $A(2)=0$ (semicircle). 

This is further corroborated in Fig. \ref{fig:twocriticallines} where we plot the same critical line $A=A(a)$ along with the line (dashed red) where a zero develops for the density at the right edge, i.e. $L_2(y_2)=0$. Moving from $A=0$ downwards at fixed $a>2$, one \emph{first} meets the line at which a zero of the density develops at the right edge, and thus where the one-cut phase with a positive density ends. Only beyond that phase boundary one meets the line where a traceless one-cut density must live. However, here the density is already no longer positive as shown in Fig. \ref{fig:critical}, because loosely speaking the zero has propagated inside the support. This convincingly corroborates the non-existence of a positive definite and traceless one-cut density anywhere in the $(A,a)$ plane, except for $A=0$ or $a\to\infty$.


\end{document}